\documentclass[twocolumn]{aastex701}
\usepackage{amsmath,amssymb}
\usepackage{graphicx}
\usepackage{natbib}
\usepackage{hyperref}
\shorttitle{Expanding bubbles in hot CGM}
\shortauthors{Li}

\begin{document}

\title{Analytical Framework for Expanding Bubbles in a Hot Circumgalactic Medium}

\correspondingauthor{Jiang-Tao Li}
\email{pandataotao@gmail.com}

\author[orcid=0000-0001-6239-3821,sname='J.-T. Li']{Jiang-Tao Li}
\affiliation{Purple Mountain Observatory, Chinese Academy of Sciences, 10 Yuanhua Road, Nanjing 210023, People’s Republic of China}
\email{pandataotao@gmail.com}  

\date{\today}

\begin{abstract}
We develop an analytic framework for the evolution of feedback-driven bubbles expanding into a hot, volume-filling circumgalactic medium (CGM), where the ambient pressure and sound speed are non-negligible and radiative cooling is often inefficient. The evolution is organized into four stages --- free expansion, Sedov--Taylor expansion, pressure-modified/transonic transition, and post-transonic relaxation --- and we derive self-consistent scalings for the characteristic radii and timescales that delimit these stages. A central result is that, in hot halos, the end of the strong-shock evolution is frequently set by pressure confinement and transonicity rather than by the onset of catastrophic cooling, implying only a modest late-time overshoot beyond the pressure-balance/transonic point. We connect the dynamics to observable outcomes by estimating bubble sizes and lifetimes, order-of-magnitude band-limited X-ray luminosities, and high-ionization ion column densities, and we provide stitched numerical trajectories that contrast our pressure-modified model appropriate for hot CGM conditions with a classical Sedov--Taylor benchmark. We then discuss physically motivated extensions beyond the single-event baseline, including continuous or episodic energy injection relevant for AGN-driven bubbles and nuclear outflows, highlighting the much higher specific energy of AGN feedback compared to supernovae and the resulting dynamical differences in how bubbles are driven. We further outline how multiphase interaction, mass loading, anisotropic dissipation, intermittency, confinement, and non-thermal channels can increase the emergent X-ray radiation efficiency without requiring changes to the intrinsic feedback energy partition at the launching site. This framework provides a transparent bridge between idealized bubble theory and feedback signatures in hot galactic halos.
\end{abstract}


\section{Introduction}\label{sec:introduction}

Feedback from supernovae (SNe), stellar wind, and active galactic nuclei (AGN) plays a central role in galaxy evolution by injecting energy, momentum, and metals into the surrounding gaseous environment. In classical models, the dynamical evolution of feedback-driven bubbles and superbubbles has been developed primarily for cold or warm interstellar media (ISM), where strong shocks, thin swept-up shells, and radiative cooling dominate the late-time evolution \citep[e.g.,][]{Castor1975,Weaver1977,Chevalier1974,Truelove1999,Cioffi1988}. These models underpin much of our intuition about how stellar and AGN feedback regulate star formation and drive large-scale outflows.

However, observations and cosmological simulations now indicate that a large fraction, if not the majority of baryons associated with galaxies reside not in the ISM but in a hot, volume-filling circumgalactic medium (CGM) with temperatures $T \sim 10^{6}$--$10^{7}\,\mathrm{K}$ \citep[e.g.,][]{Bregman2018,Bregman2022,LiJ2017,LiJ2018}. In this regime, the fundamental physical assumptions underlying classical bubble evolution theory break down. The ambient sound speed and thermal pressure are high, radiative cooling is often inefficient, and the swept-up medium carries substantial thermal energy. As a result, feedback-driven disturbances in the hot CGM can become weak shock or transonic before radiative shell formation occurs, fundamentally altering their dynamical evolution.

The importance of hot ambient media has been recognized in several specific contexts. Studies of Type~Ia supernova remnants (SNRs) expanding into tenuous, hot halos have shown that, once the expansion becomes transonic, energy is transported primarily by weak shocks and sound waves rather than by classical strong shocks, and radiative cooling remains dynamically unimportant because the cooling time exceeds the expansion and sound-crossing times \citep{Tang2005,Tang2009}. In galaxy clusters, AGN-inflated bubbles and cavities exhibit weak shock fronts, buoyant rise, and acoustic heating, highlighting the role of finite ambient pressure and sound speed in regulating feedback in hot atmospheres \citep[e.g.,][]{Churazov2001,Fabian2003,Forman2007}. Nevertheless, much of the existing analytical work on superbubbles and AGN-driven bubbles either neglects the hot-medium limit or concentrates on isolated evolutionary phases—such as the initial supersonic inflation powered by ongoing energy injection \citep[e.g.,][]{Weaver1977,MacLow1988,Begelman1989,Faucher2012} or the subsequent buoyant evolution after inflation has ceased \citep[e.g.,][]{Churazov2001,Bruggen2002}—without providing a unified evolutionary framework that connects the full sequence from supersonic expansion to pressure-regulated, transonic relaxation.

A key unresolved issue is how to connect the early, strong-shock expansion of a feedback-driven bubble to its late-time fate in a hot CGM. In particular, it remains unclear under what conditions radiative cooling produces a classical pressure-driven snowplow, and under what conditions the evolution instead terminates through pressure confinement and a transition to transonic or subsonic expansion. This distinction has direct implications for the maximum spatial reach of feedback, the lifetime of coherent bubbles, and the observability of X-ray cavities and high-ionization absorption signatures.

In this paper, we develop an analytical framework for expanding bubble evolution in a hot CGM environment that explicitly incorporates finite ambient pressure and sound speed. We focus on the controlled case of an instantaneous energy injection, which serves as a baseline for more complex continuous or episodic feedback scenarios. The evolution is organized into a sequence of physically motivated stages: (I) free expansion, (II) Sedov--Taylor expansion, (III) pressure confinement and/or transonic transition, and (IV) post-transonic evolution. A central emphasis of this work is that, in hot halos, the termination of the strong-shock phase is often set by transonicity or pressure balance rather than by radiative cooling.

We derive analytic scalings for characteristic radii, timescales, X-ray luminosities, and ionic column densities, and we compare a classical Sedov--Taylor benchmark to a fiducial pressure-modified model that captures the essential effects of hot-CGM confinement. Throughout, we aim to distinguish robust kinematic regulators—set primarily by the global CGM properties—from observables that depend sensitively on uncertain microphysics such as mixing and mass loading. This framework provides a physically transparent foundation for interpreting feedback signatures in hot galactic halos and for extending analytic models beyond the cold-ISM paradigm. 

The remaining parts of the paper are organized as follows. In \S\ref{sec:baselinemodel} we present the baseline model for a single, instantaneous SN-like energy injection expanding into a uniform hot CGM, derive the stage-by-stage evolution, and establish the role of finite ambient pressure and sound speed in regulating the pressure-modified/transonic transition. In \S\ref{sec:global_outcomes} we translate the framework into global observable outcomes, including characteristic bubble sizes and lifetimes, order-of-magnitude X-ray luminosities, and ionic column densities. In \S\ref{sec:discussion} we discuss extensions beyond the idealized baseline and provide qualitative connections to observations; in particular, \S\ref{subsec:AGN_continuous} considers continuous or episodic energy injection relevant to AGN-driven bubbles, and \S\ref{subsec:xray_enhancement} discusses how multiphase interaction, mass loading, intermittency, confinement, and non-thermal channels can enhance the emergent X-ray radiation efficiency. We summarize the main conclusions and implications in \S\ref{sec:summary}.

%===============================================================
%===============================================================
%===============================================================

% ============================================================
% Section: Single-epoch SN-like injection in a uniform hot CGM
% ============================================================

\section{Baseline Model for Expanding Bubble Evolution in a Uniform Hot CGM}
\label{sec:baselinemodel}

In this section we develop a baseline model for the evolution of an expanding bubble produced by a \emph{single, instantaneous} energy injection (e.g., a Type~Ia SN) expanding into a \emph{uniform} hot CGM. The purpose is twofold: (i) to establish a physically controlled dynamical framework that naturally incorporates the confining effect of the hot CGM pressure and sound speed, and (ii) to connect the dynamics to the key observables (size, lifetime, X-ray luminosity, and ionic columns) that motivate the later numerical or observational explorations.

Throughout this section we simply assume: (i) spherical symmetry; (ii) a uniform ambient medium; (iii) an optically thin plasma; and (iv) a thin swept-up shell separating the bubble interior from the ambient CGM.
Many other processes which may complicate the bubble evolution, such as gravitational stratification, ambient turbulence, multiphase structure, magnetic fields/cosmic rays (CRs), and anisotropic conduction, are not considered immediately. Radiative cooling is introduced explicitly later in this section, while continuous and/or episodic injections will be considered in later sections as a generalization of this baseline.

%===============================================================
\subsection{Physical setup and assumptions}
\label{subsec:setup}
%===============================================================

%---------------------------------------------------------------
\subsubsection{Ambient hot CGM properties}
\label{subsubsec:AmbientHotCGM}
%---------------------------------------------------------------

We assume a surrounding homogeneous, static, ideal-gas CGM characterized by constant number density $n_0$ and temperature $T_0$. The ambient pressure and mass density are:
\begin{equation}\label{eq:P0rho0}
P_0 = n_0 k_B T_0,
\qquad
\rho_0 = \mu m_p n_0,
\end{equation}
where $\mu$ is the mean molecular weight, $m_p$ is the proton mass, and $k_B$ is the Boltzmann constant. The adiabatic index is taken to be $\gamma=5/3$. The ambient sound speed is thus:
\begin{equation}\label{eq:cs0}
c_{s,0} = \sqrt{\frac{\gamma k_B T_0}{\mu m_p}} \simeq 151\,{\rm km\,s^{-1}}
\left(\frac{T_0}{10^6\,{\rm K}}\right)^{1/2}\left(\frac{\mu}{0.61}\right)^{-1/2},
\end{equation}
where we adopt a fiducial hot CGM temperature of $T_0=10^6\rm~K$ and mean molecular weight $\mu=0.61$.

At $t=0$, an energy $E_0$ is deposited at the origin. The resulting blast wave expands nearly spherically, sweeping ambient CGM gas into a thin shell at radius $R(t)$ with velocity $v(t)=\dot R(t)$. The swept-up shell mass is:
\begin{equation}\label{eq:Msw}
M_{\rm sw}(t) = \frac{4\pi}{3}\rho_0 R(t)^3.
\end{equation}
The cavity interior is filled with shocked gas at pressure $P_b(t)$, which drives the shell against the ambient pressure $P_0$.

%---------------------------------------------------------------
\subsubsection{Shell momentum and interior energy}
\label{subsubsec:ShellMomentumInteriorEnergy}
%---------------------------------------------------------------

We describe the bubble as a hot interior at pressure $P_b(t)$ driving a thin swept-up shell against the ambient pressure $P_0$. The shell momentum evolves according to:
\begin{equation}\label{eq:momentum}
\frac{d}{dt}\left(M_{\rm sw} v\right)=4\pi R^2\left(P_b - P_0\right),
\end{equation}
where the pressure difference across the shell is the net driving force. Eq.~\ref{eq:momentum} makes explicit that the hot CGM pressure can strongly suppress expansion at late times, in contrast to the commonly adopted $P_0\simeq 0$ approximation appropriate for cold, low-pressure media.

In early evolutionary stages, the inertia of the ejecta must also be taken into account, so the total mass of the expanding material is: $M_{\rm tot}(R) = M_{\rm ej} + M_{\rm sw}(R)$. Eq.~\ref{eq:momentum} is thus corrected to:
\begin{equation}\label{eq:momentumtot}
M_{\rm tot}\frac{dv}{dt} + v\frac{dM_{\rm sw}}{dt} =
4\pi R^{2}\left(P_{\rm b} - P_{0}\right).
\end{equation}

The thermal energy of the bubble interior is
\begin{equation}\label{eq:Eb_def}
E_b = \frac{P_b V}{\gamma-1},
\qquad
V=\frac{4\pi}{3}R^3,
\end{equation}
and evolves as
\begin{equation}\label{eq:energy_general}
\frac{dE_b}{dt} = - P_b\frac{dV}{dt} - L_{\rm cool} \equiv - P_b\,4\pi R^2 v - L_{\rm cool}.
\end{equation}
The first term on the right-hand side is the $PdV$ work done by the interior gas on the shell, while $L_{\rm cool}$ represents radiative energy losses from the hot interior (and, where relevant, the cooling layer adjacent to the shell). In the adiabatic limit $L_{\rm cool}\rightarrow 0$, Eqs.~\ref{eq:momentum} -- \ref{eq:energy_general} define a closed dynamical system once a prescription for the interior mass (hence density) is specified.

For later convenience, we define the shock Mach number
\begin{equation}
\mathcal{M}(t) \equiv \frac{v(t)}{c_{s,0}}.
\label{eq:mach_def}
\end{equation}
In a hot CGM, $c_{s,0}$ is large enough that $\mathcal{M}$ can reach unity at relatively small radii, making the transition from a strong shock to a weak shock/compression wave a central element of the evolution.

%---------------------------------------------------------------
\subsubsection{Radiative cooling and X-ray emissivity}
\label{subsubsec:norm_cooling_xray}
%---------------------------------------------------------------

Both the shocked interior and the hot CGM reaches $T\sim 10^6$--$10^8$~K and are optically thin, so radiative losses and X-ray emission are dominated by collisional plasma processes (thermal bremsstrahlung plus metal-line emission). The total cooling luminosity is:
\begin{equation}\label{eq:Lcool_full}
L_{\rm cool} = \int n_e n_i \Lambda(T,Z)\, dV,
\end{equation}
where $\Lambda(T,Z)$ is the cooling function and $n_e$ and $n_i$ are the electron and ion number densities.
Similarly, the band-limited X-ray luminosity in a chosen bandpass is:
\begin{equation}\label{eq:Lx_full}
L_X = \int n_e n_i \Lambda_X(T,Z)\, dV,
\end{equation}
where $\Lambda_X$ is the emissivity in that band.

For analytic scaling and for mapping to numerical implementation, it is convenient to express both in terms of an emission measure:
\begin{equation}\label{eq:EMdef}
{\rm EM} \equiv \int n_e n_H\, dV \simeq \bar n_e\,\bar n_H\,V,
\end{equation}
so that (up to order-unity composition factors)
\begin{equation}\label{eq:Lcool_Lx_EM}
L_{\rm cool} \simeq {\rm EM}\,\Lambda(T_b,Z),
\qquad
L_X \simeq {\rm EM}\,\Lambda_X(T_b,Z).
\end{equation}

\paragraph{Interior density and mass loading.}
We parameterize the hot interior mass as:
\begin{equation}\label{eq:Mb_fm_norm}
M_b(t) = M_{\rm ej} + f_m\,M_{\rm sw}(t),
\end{equation}
where $f_m$ is an effective mass-loading factor describing mixing/evaporation of swept-up material into the hot phase. The corresponding volume-averaged interior density is:
\begin{equation}\label{eq:nb_norm}
n_b(t) = \frac{3M_b(t)}{4\pi \mu m_p R^3}.
\end{equation}
This parameterization allows one to explore how uncertain microphysics (mixing and evaporation) impacts $L_X$ and $t_{\rm cool}$ while keeping the dynamical framework unchanged.

\paragraph{Metallicity dependence.}
In the temperature range relevant for hot bubbles, cooling/emission generally receives contributions from (i) H/He bremsstrahlung and (ii) metal-line emission. A useful bookkeeping form is: 
\begin{equation}\label{eq:Lambda_decomp}
\Lambda(T,Z) \approx \Lambda_{\rm HHe}(T) + \left(\frac{Z}{Z_\odot}\right)\Lambda_Z(T),
\end{equation}
and similarly for $\Lambda_X$ in a certain observational band. In regimes where metal-line cooling dominates, one has approximately $L_{\rm cool}\propto Z$ and $t_{\rm cool}\propto Z^{-1}$ \citep{Sutherland1993}; when bremsstrahlung dominates, the $Z$-dependence weakens substantially. We adopt $Z=Z_\odot$ as the fiducial metallicity for normalized expressions, and explicitly keep $Z/Z_\odot$ so that other values of $Z$ can be quoted by rescaling. Another caution should be made that the metallicity of the ejecta and the swept-up ambient medium is different, so when the bubble evolves, the average metallicity quoted in the above equations should change and also depends on the mixing process.

%---------------------------------------------------------------
%\subsubsection{Cooling time and the survival of hot bubbles}
%\label{subsubsec:norm_tcool}
%---------------------------------------------------------------

The characteristic radiative cooling time of the hot interior is:
\begin{equation}\label{eq:tcool}
t_{\rm cool} \equiv \frac{E_b}{L_{\rm cool}} \simeq
\frac{\tfrac{3}{2}n_b k_B T_b V}{n_b^2 \Lambda(T_b,Z)V} =
\frac{3k_B T_b}{2 n_b \Lambda(T_b,Z)}.
\end{equation}
Comparing $t_{\rm cool}$ with the dynamical time $t_{\rm dyn}\sim R/v$ determines whether the evolution remains adiabatic or becomes radiative prior to pressure confinement.

A central expectation in a hot CGM is that the low density suppresses radiative losses, such that the strong-shock phase often terminates when the bubble becomes transonic ($\mathcal{M}\rightarrow 1$) and/or pressure-confined ($P_b\rightarrow P_0$) rather than by catastrophic cooling.
Nevertheless, cooling remains essential for predicting X-ray detectability and for assessing whether the hot cavity can persist as a coherent structure before it mixes into the ambient CGM.
In later sections we evaluate $t_{\rm cool}(t)$ along each trajectory and identify the regions of parameter space where $t_{\rm cool}\lesssim t_{\rm dyn}$, and quantify how the survivability depends on $E_0$, $n_0$, $T_0$, metallicity, and the mass-loading parameter $f_m$.

\subsection{Stage definitions and ordering in a hot medium}
\label{subsec:stages_hot_medium}

The dynamical evolution of a SNR or expanding bubble can be characterized by comparing a few physically well-defined timescales. Similar as the standard SNR evolution model (e.g., \citealt{Chevalier1974,Cioffi1988,Truelove1999}), in this work, we focus on four key timescales that control the transitions between evolutionary stages and determine their ordering. A central point is that, while the early stages are largely identical to those in the standard SNR model, the ordering of the later stages can differ qualitatively in a hot medium.

\subsubsection{Key timescales}\label{subsubsec:KeyTimescales}

The four key timescales are:

\emph{1. Swept-up (ejecta--ambient equality) time, $t_{\rm sw}$.}  
This timescale marks the end of the free-expansion (ejecta-dominated) stage and is defined by the condition that the swept-up ambient mass equals the ejecta mass, $M_{\rm sw}=M_{\rm ej}$. This condition sets the corresponding radius:
\begin{equation}\label{eq:Rsw}
R_{\rm sw} = \left(\frac{3 M_{\rm ej}}{4\pi \rho_0}\right)^{1/3},
\end{equation}
where $\rho_0 = \mu m_{\rm p} n_0$ is the ambient mass density. Numerically, this radius can be written as:
\begin{equation}\label{eq:Rswnorm}
R_{\rm sw} \simeq 0.019\,
\left(\frac{M_{\rm ej}}{1\,\mathrm{M_\odot}}\right)^{1/3}
\left(\frac{n_0}{10^{-3}\,\mathrm{cm^{-3}}}\right)^{-1/3}
\ \mathrm{kpc},
\end{equation}
where a mean molecular weight $\mu\simeq0.61$ has been assumed.

Approximating the early expansion as ballistic, the swept-up time is:
\begin{equation}\label{eq:tsw}
t_{\rm sw} \simeq \frac{R_{\rm sw}}{v_{\rm ej}},
\end{equation}
with the characteristic ejecta velocity:
\begin{equation}\label{eq:vejFE}
v_{\rm ej} = \left(\frac{2E_0}{M_{\rm ej}}\right)^{1/2}.
\end{equation}
Combining the above expressions yields: 
\begin{equation}\label{eq:tswnorm}
\begin{aligned}
t_{\rm sw} \simeq {} & 1.6 \times 10^{3}\,
\left(\frac{M_{\rm ej}}{1\,\mathrm{M_\odot}}\right)^{5/6}
\left(\frac{E_0}{10^{51}\,\mathrm{erg}}\right)^{-1/2} \\
& \times
\left(\frac{n_0}{10^{-3}\,\mathrm{cm^{-3}}}\right)^{-1/3}
\ \mathrm{yr}.
\end{aligned}
\end{equation}
This transition typically occurs at very early times and depends primarily on the ejecta mass $M_{\rm ej}$, explosion energy $E_0$, and ambient number density $n_0$.

\emph{2. Radiative cooling (shell-formation) time, $t_{\rm cool}$.}  
The radiative cooling time is defined as the epoch at which radiative energy losses in the bubble interior become dynamically important, i.e., when the cooling time becomes comparable to the expansion or sound-crossing time.  

Following Eq.~\ref{eq:tcool}, the cooling time depends on the temperature and density of the hot bubble interior. For a bubble expanding into a hot CGM, the interior gas is expected to be tenuous and highly ionized, with characteristic values $T_b \sim 10^{7}$--$10^{8}\,\mathrm{K}$ and $n_b \sim 10^{-4}$--$10^{-3}\,\mathrm{cm^{-3}}$. Adopting a representative temperature $T_b = 10^{7}\,\mathrm{K}$, density $n_b = 10^{-3}\,\mathrm{cm^{-3}}$, and a solar-metallicity cooling function $\Lambda(T_b,Z_\odot) \simeq 2 \times 10^{-23}\,\mathrm{erg\,cm^{3}\,s^{-1}}$ appropriate for collisionally ionized gas \citep{Sutherland1993}, the cooling time can be written as:
\begin{equation}\label{eq:tcoolnorm}
\begin{aligned}
t_{\rm cool} \simeq {} & 3.3 \times 10^{9}\,
\left(\frac{T_b}{10^{7}\,\mathrm{K}}\right)
\left(\frac{n_b}{10^{-3}\,\mathrm{cm^{-3}}}\right)^{-1} \\
& \times
\left(\frac{\Lambda(T_b,Z)}{2 \times 10^{-23}\,\mathrm{erg\,cm^{3}\,s^{-1}}}\right)^{-1}
\ \mathrm{yr}.
\end{aligned}
\end{equation}

This timescale is orders of magnitude longer than the characteristic expansion, pressure-equilibration, or transonic timescales in the hot CGM (see below), implying that radiative cooling of the bubble interior does not lead to shell formation or a classical snowplow phase under the conditions considered here. In contrast, in cold or warm ISM where densities are much higher and cooling is more efficient, $t_{\rm cool}$ (typically $\sim10^{4-5}\rm~yr$) can become comparable to the expansion time and marks the transition from the Sedov--Taylor phase to a radiative snowplow \citep{Cioffi1988,Truelove1999}.

\emph{3. Pressure-balance time, $t_{P}$.}  
The pressure-balance time is defined by the condition that the mean interior pressure of the bubble becomes comparable to the ambient pressure,
\begin{equation}
P_{\rm b}(t_{P}) \simeq P_0 = n_0 k_{\rm B} T_0 .
\end{equation}
During the adiabatic expansion stage, the interior pressure of the bubble follows the Sedov--Taylor (ST) scaling,
\begin{equation}\label{eq:Pbttimescale}
P_{\rm b}(t) \sim \eta_{\rm th}\,\frac{E_0}{2\pi R_{\rm ST}^3(t)},
\end{equation}
where $\eta_{\rm th}$ is a dimensionless factor of order unity that accounts for the fraction of the explosion energy stored as thermal energy in the interior \citep[e.g.,][]{Weaver1977,Chevalier1974,Truelove1999}.  
Equating $P_{\rm b}$ with the ambient pressure yields a characteristic pressure-balance radius:
\begin{equation}\label{eq:RPnorm}
\begin{aligned}
R_{P} \simeq {} & 
\left(\frac{\eta_{\rm th} E_0}{2\pi P_0}\right)^{1/3}
\simeq
1.1\,
\eta_{\rm th}^{1/3}
\left(\frac{E_0}{10^{51}\,\mathrm{erg}}\right)^{1/3} \\
& \times
\left(\frac{n_0}{10^{-3}\,\mathrm{cm^{-3}}}\right)^{-1/3}
\left(\frac{T_0}{10^{6}\,\mathrm{K}}\right)^{-1/3}
\ \mathrm{kpc}.
\end{aligned}
\end{equation}

The corresponding pressure-balance time is obtained by requiring $R_{\rm ST}(t_{P}) = R_{P}$, where $R_{\rm ST}(t) = \xi (E_0/\rho_0)^{1/5} t^{2/5}$ is the standard ST solution \citep{Sedov1959,Taylor1950}. This gives:
\begin{equation}\label{eq:tPnorm}
\begin{aligned}
t_{P} \simeq {} &
2.6 \times 10^{6}\,
\eta_{\rm th}^{5/6}
\left(\frac{E_0}{10^{51}\,\mathrm{erg}}\right)^{1/3}
\left(\frac{n_0}{10^{-3}\,\mathrm{cm^{-3}}}\right)^{-1/3} \\
& \times
\left(\frac{T_0}{10^{6}\,\mathrm{K}}\right)^{-5/6}
\ \mathrm{yr}.
\end{aligned}
\end{equation}

Because the ambient pressure depends explicitly on both density and temperature, pressure confinement becomes particularly important in hot, high-sound-speed environments such as the CGM, and can occur well before radiative cooling becomes dynamically significant (Eq.~\ref{eq:tcoolnorm}).

For comparison, in the cold/warm ISM the ambient pressure is typically $P_0/k_{\rm B} \sim 10^{3}$--$10^{4}\,\mathrm{K\,cm^{-3}}$ (e.g., $n_0 \sim 0.1$--$1\,\mathrm{cm^{-3}}$ and $T_0 \sim 10^{4}\,\mathrm{K}$ for the warm phase). Using the same ST-based estimate above then gives $t_{P}\sim10^7\rm~yr$, significantly longer than the typical shell-formation time in the cold/warm ISM ($t_{\rm cool} \sim 10^{4}$--$10^{5}\,\mathrm{yr}$ for $n_0\sim0.1$--$1\,\mathrm{cm^{-3}}$), so radiative losses generally become important well before pressure balance is reached. The corresponding pressure-balance radius in the cold/warm ISM is $R_{P} \sim 0.2\rm~kpc$, which is significantly smaller than the hot-CGM value because the ISM pressure is higher. However, despite the smaller $R_P$, the corresponding $t_P$ remains long (typically $\gtrsim$~Myr) because the blast wave decelerates rapidly at high ambient density; in practice, for the cold/warm ISM the remnant becomes radiative well before reaching pressure balance ($t_{\rm cool} \ll t_P$), and the late evolution is better described by the radiative snowplow stages.

\emph{4. Transonic (Mach-1) time, $t_{\mathcal{M}}$.}  
The transonic time is defined by the condition that the shock (or expansion) velocity equals the ambient sound speed, $v_{\rm sh}(t_{\mathcal{M}}) = c_{s,0}$, corresponding to a Mach number $\mathcal{M}=1$. Beyond this point, the disturbance can no longer be described as a strong shock. Instead, the compression becomes weak, entropy generation across the front is small, and the flow transitions toward a weak shock or a compression wave.

Here a \emph{compression wave} refers to a finite-amplitude compressive disturbance in which the fluid variables (pressure, density, and velocity) vary smoothly over a finite spatial scale (as opposed to the discontinuous jump of a strong shock); it propagates at approximately the local sound speed and transports energy and momentum primarily through reversible compression rather than irreversible shock heating (e.g., \citealt{Landau1987}). In the transonic and subsonic regimes, the disturbance therefore propagates as an acoustic or pressure perturbation that gradually dissipates into the ambient medium.

Using the Sedov--Taylor (ST) solution:
\begin{equation}\label{eq:RST}
R_{\rm ST}(t) = \xi \left(\frac{E_0}{\rho_0}\right)^{1/5} t^{2/5},
\end{equation}
the shock velocity is: 
\begin{equation}\label{eq:vshST}
v_{\rm sh}(t) = \frac{{\rm d}R_{\rm ST}}{{\rm d}t}
= \frac{2}{5}\,\xi \left(\frac{E_0}{\rho_0}\right)^{1/5} t^{-3/5},
\end{equation}
where $\xi$ is the Sedov--Taylor dimensionless similarity constant that sets the normalization of the shock radius. It has a fixed value of $\xi \approx 1.15$ for $\gamma=5/3$ in the classical ST solution. On the other hand, $\eta_{\rm th}$ defined in Eq.~\ref{eq:Pbttimescale} denotes the fraction of the explosion energy residing as thermal energy in the bubble interior and has a value of $\eta_{\rm th} \approx 0.72$ in the ST phase, with the remainder residing in bulk kinetic motion. The two parameters encode distinct aspects of the Sedov--Taylor solution \citep[e.g.,][]{Sedov1959,Taylor1950,Truelove1999}.

We emphasize that these numerical values strictly apply only within the ideal Sedov--Taylor regime, which assumes adiabatic, self-similar expansion into a cold, pressureless medium. In later evolutionary stages or in a hot, pressurized medium, the flow is no longer strictly self-similar and the partition between thermal and kinetic energy may evolve as the bubble performs work against the ambient gas. In this work, the ST values of $\xi$ and $\eta_{\rm th}$ are therefore adopted as well-defined reference normalizations that anchor the early-time evolution, while allowing for modest deviations from the ideal ST assumptions in subsequent stages.

From Eq.~\ref{eq:vshST}, setting $v_{\rm sh}(t_{\mathcal{M}})=c_{s,0}$ gives:
\begin{equation}\label{eq:tM}
t_{\mathcal{M}}=
\left[\frac{2\xi}{5}\left(\frac{E_0}{\rho_0}\right)^{1/5}\frac{1}{c_{s,0}}\right]^{5/3},
\end{equation}
where $c_{s,0}=(\gamma k_{\rm B}T_0/\mu m_{\rm p})^{1/2}$. Normalizing to characteristic hot CGM conditions yields:
\begin{equation}\label{eq:tMnorm}
\begin{aligned}
t_{\mathcal{M}} \simeq {} &
3.0 \times 10^{6}\,
\left(\frac{\xi}{1.15}\right)^{5/3}
\left(\frac{E_0}{10^{51}\,\mathrm{erg}}\right)^{1/3} \\
& \times
\left(\frac{n_0}{10^{-3}\,\mathrm{cm^{-3}}}\right)^{-1/3}
\left(\frac{T_0}{10^{6}\,\mathrm{K}}\right)^{-5/6}
\ \mathrm{yr},
\end{aligned}
\end{equation}
where $\gamma=5/3$ and $\mu\simeq0.61$ have been assumed.

$t_{\mathcal{M}}$ depends on the dimensionless parameter $\xi$. The ST value of $\xi \approx 1.15$ is fixed for a given adiabatic index $\gamma=5/3$ and primarily controls the normalization of the shock radius and expansion velocity, so the time $t_{\mathcal M}$ at which the expansion velocity drops to the ambient sound speed. In contrast, $t_P$ also depends on a dimensionless parameter $\eta_{\rm th}$ (Eq.~\ref{eq:tPnorm}), which controls the normalization of the interior thermal pressure and therefore sets when the pressure equilibrium is reached. While $\xi$ should be regarded as a reference constant inherited from the ST phase, $\eta_{\rm th}$ may be interpreted as an effective thermalization parameter that can deviate modestly from its ST value (e.g., $\eta_{\rm th}\sim0.6$--0.8) once the evolution departs from ideal self-similarity. 

Regardless the specific values of $\eta_{\rm th}$ and $\xi$, the transonic time $t_{\mathcal{M}}$ is often comparable to the pressure-balance time $t_P$ in a hot media. Physically, this is because the Mach-1 condition $v_{\rm sh}\sim c_{s,0}$ implies a post-shock pressure contrast of order unity, and using the ST scalings one finds $v_{\rm sh}^2 \sim E_0/(\rho_0 R_{\rm ST}^3)$, while $c_{s,0}^2 \sim P_0/\rho_0$. Thus $v_{\rm sh}\sim c_{s,0}$ is nearly equivalent (up to order-unity constants such as $\xi$ and $\eta_{\rm th}$) to $E_0/R_{\rm ST}^3 \sim P_0$, i.e., the pressure-balance condition defining $t_P$. In the following discussions, we will consider scenarios with both $t_P>t_{\mathcal{M}}$ and $t_P<t_{\mathcal{M}}$.

For comparison, applying Eq.~\ref{eq:tMnorm} to the cold/warm ISM with characteristic warm-phase conditions ($n_0\sim 1\,\mathrm{cm^{-3}}$, $T_0\sim 10^{4}\,\mathrm{K}$) yields a transonic time of order $t_{\mathcal{M}} \sim \mathrm{few}\times10^{6}\,\mathrm{yr}$, roughly in the same order of the hot CGM case. In practice, however, SNRs in the cold/warm ISM typically become radiative much earlier ($t_{\rm cool}\sim 10^{4}$--$10^{5}\,\mathrm{yr}$ for $n_0\sim 0.1$--$1\,\mathrm{cm^{-3}}$), so radiative losses dominate the late evolution well before the shock approaches the transonic regime. As a result, the subsequent evolution is governed by radiative snowplow stages rather than by an adiabatic Sedov--Taylor approach to $\mathcal{M}\simeq1$ \citep{Cioffi1988,Truelove1999}.

\paragraph{Comparing the four timescales:}
In a cold or warm ISM (e.g., $n_0 \sim 0.1$--$10~\mathrm{cm^{-3}}$, $T_0 \sim 10^{4}$~K), the characteristic ordering of the swept-up ($t_{\rm sw}$; Eq.~\ref{eq:tswnorm}), radiative cooling ($t_{\rm cool}$; Eq.~\ref{eq:tcoolnorm}), pressure balance ($t_{P}$; Eq.~\ref{eq:tPnorm}), and transonic ($t_{\mathcal{M}}$; Eq.~\ref{eq:tMnorm}) timescales is:
\begin{equation}
t_{\rm sw} \ll t_{\rm cool} \ll t_{P} \sim t_{\mathcal{M}} .
\end{equation}
Radiative cooling therefore becomes dynamically important while the remnant is still strongly supersonic and expanding.
This leads to the canonical evolutionary sequence
\[
\begin{aligned}
& \parbox[t]{0.93\linewidth}{
Free Expansion (Ejecta-Dominated, ED) Phase
} \\
\rightarrow\ & \parbox[t]{0.93\linewidth}{
Sedov--Taylor (ST; Adiabatic Blast Wave) Phase
} \\
\rightarrow\ & \parbox[t]{0.93\linewidth}{
Radiative (Shell Formation / Pressure-Driven Snowplow, PDS) Phase
} \\
\rightarrow\ & \parbox[t]{0.93\linewidth}{
Momentum-Conserving Snowplow (MCS) Phase
} \\
\rightarrow\ & \parbox[t]{0.93\linewidth}{
Merger / Dissolution Phase
}
\end{aligned}
\]
as described in standard SNR models \citep{Chevalier1974,Truelove1999,Cioffi1988}.
In this regime, the late-time evolution is regulated primarily by radiative energy losses and momentum conservation, while ambient pressure and sound speed play only a secondary role.

In contrast, for a hot CGM (e.g., $n_0 \sim 10^{-4}$--$10^{-3}~\mathrm{cm^{-3}}$, $T_0 \sim 10^{6}$--$10^{7}$~K), the ordering of the timescales changes qualitatively. The low ambient density and the reduced cooling efficiency at temperatures above the peak of the radiative cooling curve \citep{Sutherland1993} suppress radiative losses, pushing $t_{\rm cool}$ to very late times, while the high sound speed and finite ambient pressure reduce both the transonic time $t_{\mathcal{M}}$ and the pressure-balance time $t_{P}$. For fiducial Type~Ia SN driven bubble and CGM parameters ($E_0=10^{51}$~erg, $n_0=10^{-3}~\mathrm{cm^{-3}}$, $T_0\sim10^{6}$--$2\times10^{6}$~K), one finds:
\begin{equation}
t_{\rm sw} \ll t_{P} \sim t_{\mathcal{M}} \ll t_{\rm cool},
\end{equation}
with $t_{P}$ and $t_{\mathcal{M}}$ of order a few Myr, while $t_{\rm cool}$ typically exceeds $\sim10^{9}$~yr for the hot, tenuous bubble interior.

As a result, a bubble expanding into a hot CGM can reach pressure equilibrium with its surroundings and become transonic well before radiative shell formation occurs \citep[e.g.,][]{Tang2005,Tang2009}. In this regime, the late evolution is governed primarily by pressure confinement and transonic relaxation, rather than by catastrophic radiative cooling, and the disturbance gradually dissipates into the ambient medium through weak shocks or compression waves.

\subsubsection{Stage ordering adopted in the hot CGM}
\label{subsubsec:StageOrdering}

Based on the timescale ordering discussed above, we define the evolutionary stages of a bubble expanding into a hot CGM as follows:

\begin{enumerate}

\item \textbf{Stage I: Free expansion}, $t < t_{\rm sw}$, during which the dynamics are dominated by the ejecta inertia and the swept-up ambient mass remains negligible.

\item \textbf{Stage II: Sedov--Taylor expansion}, $t_{\rm sw} < t < \min(t_{P}, t_{\mathcal{M}})$, characterized by an approximately adiabatic, self-similar blast wave in which radiative losses, ambient pressure, and finite sound speed have negligible influence on the dynamics.

\item \textbf{Stage III: Pressure-modified or transonic transition}, $\min(t_{P}, t_{\mathcal{M}}) \lesssim t \lesssim \max(t_{P}, t_{\mathcal{M}})$. This stage begins when the classical Sedov--Taylor assumptions first break down, either because the interior pressure approaches the ambient CGM pressure ($t \sim t_{P}$) or because the forward shock weakens toward Mach unity ($t \sim t_{\mathcal{M}}$). As discussed in \S\ref{subsubsec:KeyTimescales}, $t_{P}$ and $t_{\mathcal{M}}$ are typically comparable in a hot CGM but correspond to distinct physical processes. Depending on their relative ordering, Stage~III may be subdivided into two limiting regimes:
\begin{itemize}
\item \textbf{Stage IIIa ($t_{P} < t_{\mathcal{M}}$):} the interior pressure equilibrates with the ambient CGM while the expansion remains weakly supersonic. The forward shock persists but is progressively modified by the finite external pressure.
\item \textbf{Stage IIIb ($t_{\mathcal{M}} < t_{P}$):} the expansion becomes transonic before full pressure equilibrium is reached. In this case, the strong shock degenerates into a weak shock or compression wave, while a modest pressure contrast with the ambient medium may persist temporarily.
\end{itemize}
In both regimes, the classical Sedov--Taylor similarity solution ceases to apply, and the evolution becomes regulated primarily by the finite ambient pressure and sound speed, while remaining adiabatic.

If radiative cooling were to become important before either transition ($t_{\rm cool} < t_{P}$ or $t_{\mathcal{M}}$; although unlikely in a hot CGM as discussed in \S\ref{subsubsec:KeyTimescales}), the evolution would instead follow the standard cold/warm ISM pathway, with shell formation and subsequent pressure-driven or momentum-conserving snowplow stages.

\item \textbf{Stage IV: Post-equilibrium, subsonic relaxation}, $t \gtrsim \max(t_{P}, t_{\mathcal{M}})$. By the onset of this stage, the bubble has reached approximate pressure equilibrium with the ambient CGM, or may become mildly under-pressured owing to continued expansion driven by residual momentum. The expansion is fully subsonic, and the shock structure is no longer well defined, degenerating into weak compression waves or sound waves. The subsequent evolution is best described as a momentum-dominated, quasi-static relaxation of the disturbance through pressure adjustment, acoustic propagation, and mixing with the background medium, rather than as a distinct pressure-driven expansion phase.

For the hot, low-density CGM conditions considered here, radiative cooling during Stage~IV remains dynamically unimportant, as the cooling time is much longer than both the expansion and sound-crossing times and does not lead to the formation of a radiative shell. The termination of Stage~IV, and thus of the bubble expansion, may occur in one of two limiting ways. In the first, radiative cooling eventually becomes important at very late times ($t \sim t_{\rm cool}$), leading to thermal dissipation of the bubble interior. In the second, and more generic case for a hot CGM, the residual momentum is gradually exhausted, the expansion asymptotically stalls, and the disturbance fully merges with the ambient medium, leaving no dynamically distinct bubble.

\end{enumerate}

%===============================================================
\subsection{Stage I: Free expansion (ejecta-dominated)}
\label{subsec:StageI_ED}
%===============================================================

At the earliest times, the ejecta expands nearly freely, the swept-up ambient medium mass is negligible ($M_{\rm sw}\ll M_{\rm ej}$), and the dynamics are dominated by the inertia of the ejecta. To leading order the ejecta expand ballistically with an approximately constant characteristic velocity $v_{\rm ej}$, so the forward shock (or outer ejecta front) radius grows linearly with time,
\begin{equation}\label{eq:StageIRtvt}
R_{\rm sh}(t) \simeq v_{\rm ej}\, t,
\qquad
v_{\rm sh}(t) \equiv \frac{{\rm d}R_{\rm sh}}{{\rm d}t} \simeq v_{\rm ej},
\end{equation}
where the expanding velocity of the ejecta $v_{\rm ej}$ is taken to follow Eq.~\ref{eq:vejFE}.

This ``ejecta-dominated'' (ED) approximation is the standard early-time limit of SNR evolution in the ISM \citep[e.g.,][]{Chevalier1982,Truelove1999} and is equally applicable in a hot CGM prior to significant mass loading.

The ambient sound speed is approximately constant, $c_{s,0}=(\gamma k_{\rm B}T_0/\mu m_{\rm p})^{1/2}$, hence the instantaneous Mach number during Stage~I is:
\begin{equation}\label{eq:StageIMacht}
\mathcal{M}(t) \equiv \frac{v_{\rm sh}(t)}{c_{s,0}}
\simeq \frac{v_{\rm ej}}{c_{s,0}},
\end{equation}
i.e., essentially constant in time in this stage.

The shocked ambient density immediately behind the forward shock can be approximated by the strong-shock Rankine--Hugoniot jump condition \citep{Landau1987,Draine2011}:
\begin{equation}\label{eq:nshEDstrongshock}
n_{\rm sh}(t) \simeq \frac{\gamma+1}{\gamma-1}\,n_0 \simeq 4\,n_0
\qquad (\gamma=5/3,\ \mathcal{M}\gg 1),
\end{equation}
while the post-shock temperature is:
\begin{equation}\label{eq:TshEDstrongshock}
T_{\rm sh}(t) \simeq \frac{3\gamma}{16}\,\mathcal{M}^2\,T_0
\simeq \frac{5}{16}\,\mathcal{M}^2\,T_0
\qquad (\gamma=5/3,\ \mathcal{M}\gg1),
\end{equation}
which could also be expressed in the shock velocity as:
\begin{equation}\label{eq:TshvejEDstrongshock}
T_{\rm sh}(t) \simeq \frac{3}{16}\,\frac{\mu m_{\rm p}}{k_{\rm B}}\,v_{\rm sh}^2
\simeq \frac{3}{16}\,\frac{\mu m_{\rm p}}{k_{\rm B}}\,v_{\rm ej}^2.
\end{equation}
Eq.~\ref{eq:TshvejEDstrongshock} tells us that the post-shock temperature only depends on the shock velocity, and in general not on the ambient medium properties (except for the dependence on $\mu$). This is reflected by the relation between $\mathcal{M}$ and $T_0$ in Eq.~\ref{eq:TshEDstrongshock}.

In a hot CGM the Mach number may be only moderately large compared to that in the cold or warm ISM; if needed, the expressions of $n_{\rm sh}$ and $T_{\rm sh}$ under strong-shock limits above may be replaced by the general Rankine--Hugoniot jump conditions as explicit functions of the Mach number $\mathcal{M}$ \citep[e.g.,][]{Landau1987,Draine2011}:
\begin{equation}\label{eq:nshRH}
n_{\rm sh}(t) =
n_0\,\frac{(\gamma+1)\mathcal{M}^2}{(\gamma-1)\mathcal{M}^2+2},
\end{equation}
\begin{equation}\label{eq:TshRH}
T_{\rm sh}(t) =
T_0\,
\frac{\left[2\gamma \mathcal{M}^2-(\gamma-1)\right]
\left[(\gamma-1)\mathcal{M}^2+2\right]}
{(\gamma+1)^2\mathcal{M}^2}.
\end{equation}
In the strong-shock limit ($\mathcal{M}\gg1$), these expressions reduce to Eqs.~\ref{eq:nshEDstrongshock} and \ref{eq:TshEDstrongshock}, which we adopt throughout this work for simplicity.  

\paragraph{Average interior temperature in Stage~I.}
In the free-expansion stage, quoting a single ``post-shock temperature'' can be misleading: the shocked plasma need not reach instantaneous thermal (electron--ion) equilibrium, and the hot interior is dominated by the ejecta mass rather than by swept-up ambient gas. Because radiative losses are inefficient under the conditions considered here, the dominant cooling mechanism is adiabatic expansion. We therefore characterize the thermal state of the bubble by an average interior temperature defined from the interior thermal energy and gas content.

Following what we have set up in \S\ref{subsec:setup} (Eq.~\ref{eq:Mb_fm_norm}), let the total (shocked) gas mass contained within the outer radius be:
\begin{equation}\label{eq:MtotMsw}
M_{\rm tot}(t) = M_{\rm ej} + f_m M_{\rm sw}(t),
\qquad
M_{\rm sw}(t) = \frac{4\pi}{3}\rho_0 R^3(t),
\end{equation}
Here we simply assume all the SN ejecta has been shocked while a fraction $f_m$ of the swept-up ambient medium has been efficiently heated to a temperature comparable to the post-shock value, acknowledging the finite thermal equilibrium timescale in the earliest stage. We also assume a fraction $\eta_{\rm th}$ of the explosion energy be in thermal form in the bubble interior,\footnote{In Stage~I the thermal fraction is not strictly constant because the reverse shock and internal structure evolve; here $\eta_{\rm th}$ is used as an effective parameter to provide a compact analytical description.}
\begin{equation}\label{eq:EthED}
E_{\rm th}(t) \simeq \eta_{\rm th} E_0
\left(\frac{R(t)}{R_{\rm sw}}\right)^{-2}
\qquad (t<t_{\rm sw}),
\end{equation}
where the factor $(R/R_{\rm sw})^{-2}$ follows from adiabatic expansion of a monatomic gas ($\gamma=5/3$) in a volume $V\propto R^3$ (so $E_{\rm th}\propto V^{1-\gamma}\propto R^{-2}$).

Eq.~\ref{eq:EthED} should be understood as an intermediate-asymptotic description of the \emph{instantaneous} thermal energy contained within the bubble interior during the free-expansion stage, once a shocked region is established. It is not intended to be extrapolated to arbitrarily early times ($t\rightarrow 0$), because the assumptions underlying a well-defined ``bubble interior'' and meaningful volume-averaged thermodynamic quantities break down. In particular, at very early times the forward shock has only formed over a very small volume, the internal structure is rapidly evolving (non-self-similar), and additional microphysics (e.g., incomplete thermalization, electron--ion non-equilibration, and finite ejecta thickness) may be important. For these reasons, the formal divergence implied by $E_{\rm th}\propto R^{-2}$ as $R\rightarrow 0$ should be regarded as unphysical and outside the domain of applicability of our Stage~I approximation.

A simple and useful sanity check on the inferred mean temperature is provided by the specific energy of the SN ejecta. The characteristic ejecta speed may be defined by:
\begin{equation}\label{eq:vej_bound}
v_{\rm ej}\equiv\left(\frac{2E_0}{M_{\rm ej}}\right)^{1/2},
\end{equation}
corresponding to a characteristic specific energy $\varepsilon_{\rm ej}\equiv E_0/M_{\rm ej}=v_{\rm ej}^2/2$. The associated ``equivalent temperature'' is:
\begin{equation}\label{eq:Tej_bound}
T_{\rm ej}\equiv \frac{\mu m_p}{3k_B}\,v_{\rm ej}^2
\simeq \frac{2\mu m_p}{3k_B}\,\frac{E_0}{M_{\rm ej}},
\end{equation}
which represents the maximum temperature scale that can be supported by the available specific energy (up to factors of order unity depending on the degree of thermalization and the partition between ions and electrons). We therefore interpret any adopted physical characterizations of the gas temperature, e.g., the Stage~I mass-weighted mean temperature $\langle T\rangle_M$ (see below Eq.~\ref{eq:Tmass_stageI}) only at times when $\langle T\rangle_M \lesssim T_{\rm ej}$. Equivalently, we define a conservative lower limit $t_{\rm min}$ for the applicability of the Stage~I thermal-state estimates by the condition $\langle T\rangle_M(t_{\rm min}) = T_{\rm ej}$; at earlier times, the formal values of $\langle T\rangle_M$ should not be regarded as physical.

A mass-weighted mean temperature of the bubble interior may then be written as:
\begin{equation}\label{eq:Tmass_stageI}
\begin{aligned}
\langle T\rangle_M(t) {} & \equiv \frac{\mu m_{\rm p}}{k_{\rm B}}
\frac{2E_{\rm th}(t)}{3M_{\rm tot}(t)} \\
& \simeq
\frac{\mu m_{\rm p}}{k_{\rm B}}
\frac{2\eta_{\rm th}E_0}{3\left[M_{\rm ej}+f_m M_{\rm sw}(t)\right]}
\left(\frac{R(t)}{R_{\rm sw}}\right)^{-2}.
\end{aligned}
\end{equation}
In the early ejecta-dominated limit $M_{\rm sw}\ll M_{\rm ej}$, this reduces to $\langle T\rangle_M \propto R^{-2}\propto t^{-2}$, i.e., the interior cools primarily by adiabatic expansion.

For X-ray emissivity and line diagnostics, an emission-measure (EM)-weighted mean temperature is often more relevant,
\begin{equation}\label{eq:TEM_stageI}
\langle T\rangle_{\rm EM}(t) \equiv
\frac{\int n_e^2 T\,{\rm d}V}{\int n_e^2\,{\rm d}V}.
\end{equation}
A simple two-component approximation captures the qualitative behavior: an ejecta-dominated interior with characteristic density $n_{\rm ej}(t)$ and temperature $T_{\rm ej}(t)$, plus a thin shocked-ambient layer with $n_{\rm sh}\simeq 4n_0$ and temperature set by the shock speed. Because $\langle T\rangle_{\rm EM}$ weights high-density regions strongly, it is generally biased toward the denser component (often the shocked ambient layer) even when the mass-weighted mean is dominated by ejecta. In Stage~I, however, the swept-up mass and emission measure remain small, and the global evolution is still governed by adiabatic expansion rather than radiative cooling.

Although Stages~I and II are both adiabatic in the thermodynamic sense (i.e., radiative losses are negligible; see also \S\ref{subsec:stageII_ST}), Stage~I is fundamentally different from the Sedov--Taylor phase in that it is \emph{not} self-similar and the internal energy partition of the system is still evolving. This distinction is crucial for understanding the thermal evolution of the bubble interior during the free-expansion stage.

In this work, the thermal energy $E_{\rm th}(t)$ denotes the \emph{instantaneous total thermal energy contained within the bubble interior}, obtained by integrating the internal energy density over the shocked region at a given time. It is therefore a global, Eulerian bookkeeping quantity, rather than the thermal energy of a fixed parcel of gas. During Stage~I, the shock structure and velocity field continue to evolve as newly shocked ejecta and ambient material are incorporated into the bubble. As a result, the explosion energy is continuously redistributed between bulk kinetic and thermal forms, and the system has not yet settled into a fixed energy partition.

Under these conditions, although new thermal energy is continuously generated at the shock, previously shocked gas also undergoes expansion and acceleration, performing $PdV$ work and contributing increasingly to the bulk kinetic energy of the flow. The net effect is that the \emph{instantaneous} thermal energy content of the bubble interior decreases with expansion, following $E_{\rm th}\propto R^{-2}$ (Eq.~\ref{eq:EthED}). Consequently, the effective thermal fraction $\eta_{\rm th}\equiv E_{\rm th}/E_0$ is time-dependent during Stage~I. This behavior does not imply radiative cooling or a loss of total energy, but rather reflects the non-self-similar, dynamically evolving nature of the flow.

Because of this evolving energy partition, the thermal state of the bubble interior in Stage~I cannot be characterized by a unique or invariant temperature or pressure profile. Instead, we describe it using a set of representative mean quantities, including the mean density $\bar n$, mean pressure $\bar P$, and characteristic temperatures defined above, such as the mass-weighted temperature $\langle T\rangle_M$ and the emission-measure--weighted temperature $\langle T\rangle_{\rm EM}$. The Stage~I interior mean density can be estimated directly from the total gas mass (Eq.~\ref{eq:MtotMsw}) and the bubble volume, $\bar n=M_{\rm tot}/V$. The mean pressure must be computed consistently from the evolving thermal energy (Eq.~\ref{eq:EthED}) via $\bar P=(\gamma-1)E_{\rm th}/V$.

In contrast, during Stage~II the flow approaches a self-similar Sedov--Taylor solution. The internal structure of the shock becomes time-invariant when expressed in terms of the similarity variable $r/R(t)$, and the partition of the explosion energy between thermal and kinetic components asymptotically approaches a constant. As a result, the interior thermal energy becomes time-independent, $E_{\rm th}=\eta_{\rm th}E_0=\mathrm{const}$, and the mean pressure and temperature can be consistently derived from the Sedov--Taylor similarity solution. This fundamental difference explains why the evolving thermal energy and stage-dependent definitions adopted in Stage~I are neither required nor appropriate in the Sedov--Taylor phase.

%===============================================================
\subsection{Stage~II: Sedov--Taylor evolution}
\label{subsec:stageII_ST}
%===============================================================

After the ejecta--swept-up transition (Stage~I), the expansion enters an energy-conserving Sedov--Taylor (ST) phase in which the swept-up ambient mass dominates the inertia. In this stage, the dynamics approach the self-similar solution for a blast wave propagating into a uniform medium, as originally derived by \citet{Sedov1959} and \citet{Taylor1950}. In the hot CGM conditions of interest here, radiative cooling remains inefficient, so the total mechanical energy of the expanding bubble may be taken as approximately conserved throughout Stage~II, analogous to the classical adiabatic phase of SNR evolution \citep[e.g.,][]{Chevalier1974,Truelove1999}.

Throughout this subsection we explicitly distinguish between (i) \emph{shock-front} (post-shock) values evaluated immediately behind the forward shock and (ii) \emph{interior-average} values defined from the total interior gas content and thermal energy, in the same spirit as our Stage~I averages.

\paragraph{ST dynamics: radius and velocity.}
For an energy-conserving blast wave expanding into a uniform medium of density $\rho_0$, the Sedov--Taylor similarity solution gives:
\begin{equation}\label{eq:ST_R_v}
\begin{aligned}
{} & R(t)=\xi \left(\frac{E_0}{\rho_0}\right)^{1/5} t^{2/5}, \\
& v(t)\equiv \dot R(t)=\frac{2}{5}\frac{R(t)}{t}
=\frac{2}{5}\,\xi \left(\frac{E_0}{\rho_0}\right)^{1/5} t^{-3/5},
\end{aligned}
\end{equation}
where $E_0$ is the conserved explosion energy defined in \S\ref{subsubsec:AmbientHotCGM} and $\xi$ is a dimensionless constant of order unity that depends weakly on the adiabatic index $\gamma$ (for $\gamma=5/3$, $\xi\simeq 1.15$; e.g., \citealt{Sedov1959,Truelove1999}). The corresponding forward-shock Mach number is therefore:
\begin{equation}\label{eq:ST_Mach}
\mathcal{M}(t) = \frac{2}{5}\,\frac{\xi}{c_{s,0}}\left(\frac{E_0}{\rho_0}\right)^{1/5} t^{-3/5}.
\end{equation}
In a hot medium, $c_{s,0}$ is large and $\mathcal{M}(t)$ declines rapidly with time, so the ST phase is inevitably finite and will terminate as the shock approaches the transonic regime, even in the absence of strong radiative losses \citep[e.g.,][]{Tang2005}.

\paragraph{Shock-front (post-shock) values.}
For a strong shock ($\mathcal{M}\gg 1$), the Rankine--Hugoniot jump conditions yield the immediate post-shock density and temperature following Eqs.~\ref{eq:nshEDstrongshock} and \ref{eq:TshEDstrongshock}, or in a weak shock condition in the later stages of the ST stage, following Eqs.~\ref{eq:nshRH} and \ref{eq:TshRH}.
Substituting Eq.~\ref{eq:ST_Mach} into these relations, we get:
\begin{equation}\label{eq:ST_rhoshPshTsh}
\begin{aligned}
\rho_{\rm sh}(t) &=
\frac{\gamma+1}{\gamma-1}\rho_0
=4\rho_0,
\qquad
n_{\rm sh}(t)=4n_0,
\\
P_{\rm sh}(t) &\simeq \frac{2}{\gamma+1}\rho_0 v^2(t)
=\frac{3}{4}\rho_0 v^2(t)
\propto t^{-6/5},
\\
T_{\rm sh}(t) &\simeq
\frac{2(\gamma-1)}{(\gamma+1)^2}\frac{\mu m_p}{k_B}v^2(t)
=
\frac{3}{16}\frac{\mu m_p}{k_B}v^2(t)
\propto t^{-6/5}.
\end{aligned}
\end{equation}
These shock-front quantities depend only on the instantaneous Mach number and provide a direct link between the ST dynamics and observable post-shock emission properties.

\paragraph{Interior-average values: density, pressure, and temperature.}
Following the mass bookkeeping introduced in Stage~I, the total shocked gas mass enclosed within the forward shock is still Eq.~\ref{eq:MtotMsw}. The corresponding interior-average number density is:
\begin{equation}\label{eq:ST_nbar}
\bar n(t)\equiv
\frac{M_{\rm tot}(t)}{\mu m_p \, (4\pi/3)R^3(t)} =
\frac{3M_{\rm ej}}{4\pi \mu m_p R^3(t)} + f_m n_0.
\end{equation}
Once the interior gas content is dominated by swept-up material, the ejecta term becomes negligible and $\bar n(t)\simeq f_m n_0$ is approximately constant in time, as expected for an adiabatic ST phase \citep[e.g.,][]{Cox1972}. This differs from the shock front value by a factor of $\sim4$, which indicates strong inhomogeneity of gas density inside the bubble interior.

To define an interior-average pressure consistent with our Stage~I energy-based temperature, similar as in Eq.~\ref{eq:EthED}, we write the interior thermal energy as a fixed fraction of the conserved energy, $E_{\rm th}(t) \equiv \eta_{\rm th}\,E_0$, where $\eta_{\rm th}$ is an order-unity constant determined by the ST similarity solution \citep[e.g.,][]{Sedov1959,Truelove1999}. The interior-average pressure then follows as:
\begin{equation}\label{eq:ST_Pbar}
\bar P(t)\equiv (\gamma-1)\frac{E_{\rm th}(t)}{V} =
(\gamma-1)\frac{3\eta_{\rm th}E_0}{4\pi R^3(t)} \propto t^{-6/5},
\end{equation}
considering the $R(t)$ expression in Eq.~\ref{eq:ST_R_v}.

The interior-average temperature is defined as: $\bar T(t)\equiv \bar P(t)/(k_B \bar n(t))$. Substituting Eq.~\eqref{eq:ST_nbar} into the definition of $\bar T(t)$, we obtain:
\begin{equation}\label{eq:ST_Tbar_general}
\bar T(t) = \frac{\bar P(t)}{k_B \bar n(t)} =
\frac{(\gamma-1)\,3\eta_{\rm th}E_0}{4\pi k_B R^3(t)}
\left[
\frac{3M_{\rm ej}}{4\pi \mu m_p R^3(t)} + f_m n_0
\right]^{-1}.
\end{equation}
Using the Sedov--Taylor radius $R(t)$ obtained in Eq.~\ref{eq:ST_R_v}, we can rewrite the above as a function of time:
\begin{equation}\label{eq:ST_Tbar_compact}
\begin{aligned}
& {} \bar T(t) =
\frac{A}{\mathcal{R}(t)}
\left[
\frac{B}{\mathcal{R}(t)} + f_m n_0
\right]^{-1}
= \frac{A}{B + f_m n_0\,\mathcal{R}(t)}, \\
& A \equiv \frac{(\gamma-1)\,3\eta_{\rm th}E_0}{4\pi k_B},
\qquad
B \equiv \frac{3M_{\rm ej}}{4\pi \mu m_p}, \\
& \mathcal{R}(t)\equiv R^3(t)
=\xi^3 \left(\frac{E_0}{\rho_0}\right)^{3/5} t^{6/5},
\end{aligned}
\end{equation}
Finally substituting $\mathcal{R}(t)=\xi^3 (E/\rho_0)^{3/5} t^{6/5}$ yields the explicit time-dependent form
\begin{equation}\label{eq:ST_Tbar_explicit}
\begin{aligned}
\bar T(t) & {} =
\frac{(4\pi k_B)^{-1}(\gamma-1)\,3\eta_{\rm th}E_0 }
{3M_{\rm ej}(4\pi \mu m_p)^{-1}
+ (\mu m_p)^{-3/5}f_m \,\xi^3 E_0^{3/5} n_0^{2/5} t^{6/5}}\\
& \propto\ \left(\mathrm{const}+t^{6/5}\right)^{-1}.
\end{aligned}
\end{equation}
In the swept-up--dominated limit ($f_m n_0\,R^3 \gg 3M_{\rm ej}(4\pi \mu m_p)^{-1}$), this reduces to the familiar Sedov--Taylor scaling:
\begin{equation}\label{eq:ST_Tbar_asymptotic}
\bar T(t) =
\frac{(\gamma-1)\,3\eta_{\rm th}}{4\pi k_B (\mu m_p)^{-3/5} f_m \,\xi^3}
\left(\frac{E_0}{n_0}\right)^{2/5} t^{-6/5}.
\end{equation}

\paragraph{Shock-front versus interior-average scalings.}
Both the shock-front (Eq.~\ref{eq:ST_rhoshPshTsh}) and interior-average pressures and temperatures (Eqs.~\ref{eq:ST_Pbar} and \ref{eq:ST_Tbar_asymptotic}) decline as $t^{-6/5}$ during Stage~II, reflecting their common dependence on either $v^2(t)$ or $R^{-3}(t)$. However, their physical meanings differ: $(n_{\rm sh},P_{\rm sh},T_{\rm sh})$ describe the freshly shocked ambient gas immediately behind the forward shock, whereas $(\bar n,\bar P,\bar T)$ characterize the bulk interior state set by the integrated mass and thermal energy. In particular, while the post-shock density remains fixed at $n_{\rm sh}=4n_0$ for a strong shock, the interior-average density asymptotes to $\bar n\simeq f_m n_0$. This distinction is crucial in a hot CGM, where the termination of Stage~II can be defined either by a transonic condition $\mathcal{M}\rightarrow 1$ (Eq.~\ref{eq:ST_Mach}) or by pressure equilibrium with the ambient medium $\bar P\rightarrow P_0$ (Eq.~\ref{eq:ST_Pbar}), rather than by radiative shell formation as in the classical ISM case \citep[e.g.,][]{Cioffi1988}.

%===============================================================
\subsection{Stage~III: pressure-modified or transonic expansion}
\label{subsec:stageIII_PModifyTransonic}
%===============================================================

In a hot CGM, the Sedov--Taylor phase described in Stage~II (\S\ref{subsec:stageII_ST}) is not expected to persist indefinitely, even when radiative cooling is dynamically negligible. Instead, the classical ST assumptions eventually break down because the ambient thermal pressure and sound speed are finite and dynamically important. We define Stage~III as the transitional regime during which the evolution departs from the classical ST similarity solution, but before the system settles into a fully subsonic, pressure-regulated continuation (Stage~IV).

Using the Stage~II solution as a reference, the onset of Stage~III may be characterized by two closely related transition conditions. The first is the \emph{pressure-equilibrium condition}, defined by the time $t_P$ (Eq.~\ref{eq:tPnorm}), at which the interior-average pressure of the bubble becomes comparable to the ambient CGM pressure $P_0$. The second is the \emph{transonic condition}, defined by the time $t_{\mathcal{M}}$ (Eq.~\ref{eq:tMnorm}), at which the forward shock weakens to Mach unity. As discussed in \S\ref{subsec:stages_hot_medium}, in a hot CGM these two timescales are typically comparable, but their relative ordering depends on the detailed values of the thermalization efficiency $\eta_{\rm th}$ (Eq.~\ref{eq:Pbttimescale}) and the dimensionless parameter $\xi$ (Eq.~\ref{eq:RST}). For clarity, we therefore distinguish two possible evolutionary paths through Stage~III (IIIa and IIIb), depending on whether pressure equilibrium or transonicity is reached first. We emphasize that the effects associated with finite ambient pressure and with shock weakening/transonicity generally occur simultaneously in practice. The subdivision of Stage~III into two sub-stages is therefore introduced purely for clarity of discussion.

Throughout Stage~III, the dynamics of the swept-up shell are governed by the same radial momentum equation introduced in \S\ref{subsec:setup} (Eq.~\ref{eq:momentum}). In contrast to Stage~II, where the dynamics are closed by strong-shock jump conditions and the assumption $\bar{P}\gg P_0$, the relevant driving pressure in Stage~III is the interior-average pressure $\bar{P}$. This quantity provides a physically and observationally meaningful characterization of the bubble interior once the shock weakens, the post-shock temperature jump becomes moderate, and the interior approaches a state of approximate thermal equilibrium.

All evolutionary stages discussed in this work may therefore be viewed as different closures of the same underlying momentum equation (Eq.~\ref{eq:momentum}). Stage~III marks the regime in which the strong-shock closure underlying the ST solution ceases to be valid, and the evolution becomes regulated by the finite ambient pressure and sound speed, while remaining adiabatic.

%---------------------------------------------------------------
\subsubsection{Stage~IIIa: pressure-modified supersonic expansion ($t_P < t_{\mathcal{M}}$)}
\label{subsubsec:StageIIIaPressureModified}
%---------------------------------------------------------------

If pressure equilibrium is approached before the forward shock becomes transonic ($t_P < t_{\mathcal{M}}$), the expansion remains supersonic and the forward shock is still well defined, but the ambient pressure term in Eq.~\ref{eq:momentum} can no longer be neglected. In this regime, the dynamics follow Eqs.~\ref{eq:momentum} and \ref{eq:energy_general}, with the radiative cooling term neglected.

Because radiative losses are negligible, the bubble interior evolves adiabatically. Combining Eqs.~\ref{eq:Eb_def} and \ref{eq:energy_general} yields:
\begin{equation}\label{eq:IIIa_PofR_general}
\bar{P}\,V^{\gamma} = \mathrm{const}
\qquad\Rightarrow\qquad
\bar{P}(R)=\bar{P}_\ast\left(\frac{R}{R_\ast}\right)^{-3\gamma},
\end{equation}
where $(R_\ast,\bar{P}_\ast,v_\ast)$ denote the values at the Stage~II$\to$III transition, with $t_\ast=t_P$ for Stage~IIIa and $v_\ast\equiv \dot{R}(t_\ast)$.

With $\bar{P}(R)$ specified and replacing $P_b$ by $\bar{P}$ in Eq.~\ref{eq:momentum} (with $M_{\rm sw}$ given by Eq.~\ref{eq:Msw}), the momentum equation admits a first integral. Writing $v=dR/dt$ and defining $y\equiv R^3 v$, we obtain for a monatomic plasma ($\gamma=5/3$, so that $\bar{P}\propto R^{-5}$):
\begin{equation}\label{eq:IIIa_first_integral}
\begin{aligned}
y^2(R) &= y_\ast^2 + \frac{6}{\rho_0}
\Bigg[\bar{P}_\ast R_\ast^{5}(R-R_\ast) - \frac{P_0}{6}(R^6-R_\ast^6) \Bigg], \\
y_\ast &\equiv R_\ast^3 v_\ast .
\end{aligned}
\end{equation}
The corresponding velocity--radius relation is:
\begin{equation}\label{eq:IIIa_v_of_R}
\begin{aligned}
& {} v(R) = \frac{y(R)}{R^3} \\
& = \frac{1}{R^3} \Bigg\{R_\ast^6 v_\ast^2
+ \frac{6}{\rho_0}\Bigg[\bar{P}_\ast R_\ast^{5}(R-R_\ast) - \frac{P_0}{6}(R^6-R_\ast^6)\Bigg] \Bigg\}^{1/2}.
\end{aligned}
\end{equation}
This provides an explicit solution in quadrature:
\begin{equation}\label{eq:IIIa_t_of_R}
t-t_\ast = \int_{R_\ast}^{R} \frac{dR'}{v(R')}.
\end{equation}
In practice, one evaluates Eq.~\ref{eq:IIIa_t_of_R} to obtain $t(R)$, numerically inverts it to obtain $R(t)$, and then substitutes $R(t)$ into Eq.~\ref{eq:IIIa_v_of_R} to obtain $v(t)$ and into Eq.~\ref{eq:IIIa_PofR_general} to obtain $\bar{P}(t)$.

\paragraph{Mean density and temperature.} To derive the interior-average number density and temperature, we use
the total shocked gas mass contained within $R$ as defined in \S\ref{subsec:setup}, $M_{\rm tot}(t)=M_{\rm ej}+f_m M_{\rm sw}(t)$ with $M_{\rm sw}=(4\pi/3)\rho_0 R^3$ (Eq.~\ref{eq:Msw}). The interior-average mass density and number density could then be expressed with $R(t)$ as:
\begin{equation}\label{eq:IIIa_nbar}
\begin{aligned}
\bar{\rho}(t) &= \frac{M_{\rm ej}+f_m (4\pi/3)\rho_0 R^3(t)}{(4\pi/3)R^3(t)}, \\
\bar{n}(t) &= \frac{\bar{\rho}(t)}{\mu m_p}.
\end{aligned}
\end{equation}
The corresponding interior-average temperature follows from the ideal-gas law:
\begin{equation}\label{eq:IIIa_Tbar}
\bar{T}(t)=\frac{\bar{P}(t)}{\bar{n}(t)\,k_B},
\end{equation}
with $\bar{P}(t)$ given by Eq.~\ref{eq:IIIa_PofR_general} after substituting $R(t)$ from Eq.~\ref{eq:IIIa_t_of_R}.

\paragraph{Asymptotic scaling for $P_0\rightarrow0$.} If one formally sets $P_0=0$ in Eq.~\ref{eq:IIIa_v_of_R}, then at $R\gg R_\ast$ the pressure term dominates and $v(R)\propto R^{-5/2}$, implying:
\begin{equation}\label{eq:IIIa_scalings_P0zero}
\begin{aligned}
& R(t) \propto (t-t_\ast)^{2/7}, \\
& v(t) \propto (t-t_\ast)^{-5/7}, \\
& \bar{P}(t) \propto (t-t_\ast)^{-10/7}, \\
& \bar{n}(t)\simeq (f_m\rho_0)/(\mu m_p)\sim \text{constant, if } f_m M_{\rm sw}\gg M_{\rm ej}, \\
& \bar{T}(t)\propto \bar{P}(t)\propto (t-t_\ast)^{-10/7} .
\end{aligned}
\end{equation}
We stress that this $P_0\to0$ scaling corresponds to the Stage~IIIa \emph{adiabatic interior} closure and does not reproduce the Sedov--Taylor similarity law. However, physically, when $P_0$ is negligible the system remains in Stage~II and the Sedov--Taylor solution applies.

It is therefore useful to clarify the physical meaning of the Stage~IIIa closure and its relation to the classical Sedov--Taylor solution. In both Stage~II and Stage~IIIa we neglect radiative cooling, so the total energy of the shocked system is approximately conserved; thermal energy is continuously converted into kinetic energy through $P\,dV$ work in both stages. The key distinction is therefore not whether energy is conserved, but how the dynamics are \emph{closed}. In Stage~II, the strong-shock and self-similarity assumptions lead to the ST solution, in which the interior structure and driving pressure are fixed by the similarity profile and the conserved energy scale, yielding $R\propto t^{2/5}$. In Stage~IIIa, by contrast, the ST similarity closure is no longer adopted because the ambient pressure term becomes dynamically relevant even while the shock may remain weakly supersonic; we therefore evolve the system using the momentum equation explicitly and close the interior state by adiabatic thermodynamics, $\bar{P}V^\gamma=\mathrm{const}$ (Eq.~\ref{eq:IIIa_PofR_general}). Stage~IIIa thus corresponds to a regime in which the bubble expansion is increasingly regulated by the ambient pressure. The net force $4\pi R^2(\bar{P}-P_0)$ reflects the residual pressure contrast, which decreases with time and therefore leads to a gradual deceleration of the expansion. Formally taking $P_0\to0$ within this Stage~IIIa closure does not reproduce the ST similarity law; it yields the scaling implied by the adiabatic interior closure (Eq.~\ref{eq:IIIa_scalings_P0zero}). Physically, when the ambient pressure is negligible the system does not enter Stage~IIIa and remains in the Sedov--Taylor regime until the shock weakens toward Mach unity.

%---------------------------------------------------------------
\subsubsection{Stage~IIIb: transonic-first evolution and weak-compression closure ($t_{\mathcal{M}} < t_P$)}
\label{subsubsec:StageIIIbTransonicWeakShock}
%---------------------------------------------------------------

If the forward shock weakens to $\mathcal{M}\sim1$ before pressure equilibrium is reached ($t_{\mathcal{M}} < t_P$), the strong-shock closure adopted in Stage~II is no longer valid. In this regime, the propagation speed of the forward disturbance is no longer determined by strong-shock jump conditions, but by the interaction between the modest interior pressure excess and the ambient sound speed.

To obtain a conservative closure in the transonic or weak-compression regime, we introduce a characteristic post-disturbance pressure $P_2$ and relate it to the interior-average pressure via:
\begin{equation}\label{eq:fP_def}
P_2(t) \equiv f_P\,\bar{P}(t),
\qquad
f_P \sim \mathcal{O}(1),
\end{equation}
where $\mathcal{O}(1)$ means $f_P$ is a dimensionless constant of order unity, encapsulating the modest difference between the local pressure associated with the forward disturbance and the volume-averaged interior pressure, without invoking strong-shock jump conditions that are no longer valid once the expansion approaches the transonic regime.

To relate the propagation speed of the forward disturbance (instead of shock) to the interior state in Stage~IIIb, we adopt a weak-compression closure that connects smoothly to the marginally supersonic limit. Specifically, we use the standard pressure--Mach number relation for a normal weak disturbance: 
\begin{equation}\label{eq:weakshock_jump}
\frac{P_2}{P_0} = 1 + \frac{2\gamma}{\gamma+1}(\mathcal{M}^2-1),
\end{equation}
but interpret it here as an effective transonic mapping between the pressure contrast and the propagation speed, rather than as a literal discontinuous shock jump. With $P_2=f_P\bar{P}$ (Eq.~\ref{eq:fP_def}), this yields:
\begin{equation}\label{eq:IIIb_M_of_P}
\mathcal{M}(t)=
\left[1+\frac{\gamma+1}{2\gamma}\left(\frac{f_P \bar{P}(t)}{P_0}-1\right)\right]^{1/2},
\end{equation}
and hence:
\begin{equation}\label{eq:branchB_velocity}
\dot{R}(t)=c_{s,0}\,\mathcal{M}(t).
\end{equation}
This closure is intended to be used only in the near-transonic regime, where the pressure contrast is modest and $P_2$, $\bar{P}$, and $P_0$ are all within factors of a few. As the disturbance becomes fully subsonic and increasingly acoustic, the notion of a well-defined $P_2$ loses meaning and the evolution is better described by the subsonic relaxation framework of Stage~IV.

Throughout Stage~IIIb, the bubble interior continues to evolve adiabatically, so that $\bar{P}(R)$ follows the same relation as in Stage~IIIa (Eq.~\ref{eq:IIIa_PofR_general}; but here $R_\ast$ and $\bar{P}_\ast$ denote values for Stage IIIb, i.e., at $t_\ast=t_{\mathcal{M}}$). Substituting this relation into Eqs.~\ref{eq:IIIb_M_of_P} and \ref{eq:branchB_velocity} yields a closed first-order equation for the radial evolution: 
\begin{equation}\label{eq:IIIb_ODE}
\dot{R}(t) = c_{s,0}
\left[1+\frac{\gamma+1}{2\gamma}
\left(f_P\frac{\bar{P}_\ast}{P_0}
\left(\frac{R}{R_\ast}\right)^{-5} - 1 \right)
\right]^{1/2},
\end{equation}
which can be used to numerically derive $R(t)$ and then the time evolution of other physical parameters [$\mathcal{M}(t)$, $v(t)$, $\bar{P}(t)$, $\bar{n}(t)$, $\bar{T}(t)$] using Eqs.~\ref{eq:IIIa_nbar}, \ref{eq:IIIa_Tbar}, and \ref{eq:fP_def}-\ref{eq:branchB_velocity}. 

In the formal limit $f_P\bar{P}\gg P_0$, this equation reduces to the same pressure-dominated asymptotic behavior derived for Stage~IIIa (Eq.~\ref{eq:IIIa_scalings_P0zero}). In practice, however, the evolution in Stage~IIIb departs from a simple power law once the disturbance becomes fully transonic and approaches the subsonic relaxation phase described in Stage~IV.

%===============================================================
\subsection{Stage~IV: post-equilibrium subsonic relaxation}
\label{subsec:stageIV_subsonic}
%===============================================================

Stage~IV describes the late-time evolution after the system has reached the end of the Stage~III transition regime. We define the onset of Stage~IV at:
\begin{equation}
t_{\rm IV} \equiv \max(t_P,\,t_{\mathcal{M}}),
\end{equation}
so that by construction the bubble is both close to pressure balance and no longer supersonic. In this regime, the forward disturbance degenerates into weak compression waves or sound waves and a sharp shock front is no longer well defined. Nevertheless, the global expansion can continue for some time because the swept-up shell retains a finite outward momentum at $t=t_{\rm IV}$.

The key physical point is that, once the bubble enters Stage~IV, the interior pressure evolution continues to follow the adiabatic relation already obtained in Stage~III (Eq.~\ref{eq:IIIa_PofR_general}), while the net driving force in the momentum equation (Eq.~\ref{eq:momentum}) becomes small and can change sign. In particular, if the bubble continues to expand beyond $R(t_{\rm IV})$, then $\bar{P}(R)$ decreases rapidly with $R$ and the pressure contrast $(\bar{P}-P_0)$ generically becomes negative. The ambient medium therefore acts as a restoring force that decelerates the expansion.

\subsubsection{Momentum-limited overshoot and stall}
\label{subsubsec:stageIV_stall}

Let $R_{\rm IV}\equiv R(t_{\rm IV})$ and $v_{\rm IV}\equiv v(t_{\rm IV})$ denote the radius and expansion speed at the onset of Stage~IV. The subsequent evolution may be viewed as an inertial overshoot past approximate pressure equilibrium followed by ambient-pressure braking. A convenient diagnostic is the variable $y\equiv R^3 v$ introduced in Stage~IIIa, for which the first-integral form (Eq.~\ref{eq:IIIa_first_integral}) directly provides the velocity--radius relation through Stage~IV as long as the interior remains adiabatic.

We define the \emph{stall radius} $R_{\rm stall}$ as the maximum radius reached before the expansion halts:
\begin{equation}
v(R_{\rm stall}) = 0
\qquad \Leftrightarrow \qquad
y(R_{\rm stall}) = 0.
\label{eq:Rstall_def}
\end{equation}
Because $\bar{P}(R)$ declines more rapidly with radius than the ambient pressure, Eq.~\ref{eq:Rstall_def} admits a finite solution $R_{\rm stall}>R_{\rm IV}$ in the generic hot-CGM case. Physically, $R_{\rm stall}$ quantifies the maximum spatial reach of the disturbance once the strong-shock phase has terminated through pressure confinement and/or transonicity.

The corresponding stall time is obtained by integrating the inverse velocity along the trajectory,
\begin{equation}
t_{\rm stall}-t_{\rm IV} = \int_{R_{\rm IV}}^{R_{\rm stall}} \frac{dR}{v(R)},
\label{eq:tstall_def}
\end{equation}
where $v(R)$ is given by the Stage~IIIa first-integral mapping (Eqs.~\ref{eq:IIIa_first_integral}--\ref{eq:IIIa_v_of_R}) evaluated with the Stage~IV initial condition $(R_{\rm IV},v_{\rm IV},\bar{P}_{\rm IV})$.
In practice, $R_{\rm stall}$ can be found by solving $y^2(R)=0$ and then substituting into Eq.~\ref{eq:tstall_def}.

After $t_{\rm stall}$, the bubble no longer expands as a dynamically distinct structure; the residual disturbance propagates primarily as sound waves and is expected to dissipate and mix into the ambient CGM on a sound-crossing timescale of order $R_{\rm stall}/c_{s,0}$, with details depending on turbulence, conduction, and other microphysics not included in the baseline model.

As will later be presented in \S\ref{subsec:numerical_trajectories}, the late-time behavior of the expansion velocity differs qualitatively from the results of \citet{Tang2005}, who found that the velocity asymptotically approaches the ambient sound speed. The apparent discrepancy arises primarily from the different definitions of the velocity being considered. In \citet{Tang2005}, the quoted velocity corresponds to the propagation speed of the outer compressive blast wave (i.e., the shock or compression front) through the ambient medium. As the shock weakens in a hot medium, this disturbance naturally approaches the sound speed and continues to propagate outward as a weak compression wave. In contrast, the velocity discussed in this work refers to the bulk motion of the expanding bubble material (or the swept-up shell) that defines the dynamically coherent bubble structure. In our framework the stall time $t_{\rm stall}$ marks the point when the net pressure force becomes insufficient to sustain further bulk expansion of this structure. After this time, the coherent bubble expansion effectively ceases, while the residual disturbance propagates primarily as sound waves or weak compression waves in the ambient medium. Therefore, the asymptotic behavior found by \citet{Tang2005} most likely corresponds to the propagation of this residual compressive disturbance rather than the continued expansion of the bubble material itself.

\subsubsection{Cooling-dominated termination}
\label{subsubsec:stageIV_cooling}

A second, formally possible termination channel is that radiative losses eventually become dynamically important. This occurs if the interior cooling time (Eq.~\ref{eq:tcool}), evaluated along the Stage~IV trajectory using the same mass-loading prescription (Eq.~\ref{eq:Mb_fm_norm}) and interior thermodynamic variables, becomes comparable to or shorter than the characteristic evolution timescale (e.g., $R/v$ or the sound-crossing time). In that case, the adiabatic closure implicit in Eq.~\ref{eq:IIIa_PofR_general} would break down and cooling would accelerate the decay of the pressure support.

For the hot, low-density CGM conditions emphasized in this work, however, $t_{\rm cool}$ is typically orders of magnitude longer than the Myr-scale dynamical times relevant for Stages~II--IV (\S\ref{subsubsec:KeyTimescales}; also supported by observational results, e.g., \citealt{LiJ2011,LiJ2013a,LiJ2017}), so cooling is unlikely to control the end of the evolution. We therefore treat momentum exhaustion and the attainment of $R_{\rm stall}$ as the primary termination mechanism in Stage~IV.

\subsubsection{Approximate expressions for $R_{\rm stall}$ and $t_{\rm stall}$}
\label{subsubsec:stageIV_Rstall_tstall}

\paragraph{Exact stall condition.}
Using the Stage~IIIa first-integral relation between $y\equiv R^3 v$ and $R$ (Eq.~\ref{eq:IIIa_first_integral}), evaluated with $\ast\rightarrow{\rm IV}$, the stall condition $y(R_{\rm stall})=0$ yields the exact algebraic equation:
\begin{equation}\label{eq:stageIV_Rstall_exact}
\begin{aligned}
& P_0 R_{\rm stall}^{6}-6\,\bar P_{\rm IV}\,R_{\rm IV}^{5}\,R_{\rm stall} \\
& +\left(6\,\bar P_{\rm IV}\,R_{\rm IV}^{6}
-P_0 R_{\rm IV}^{6}
-\rho_0 R_{\rm IV}^{6} v_{\rm IV}^2
\right)=0 .
\end{aligned}
\end{equation}

Introducing the dimensionless variables:
\begin{equation}
x \equiv \frac{R_{\rm stall}}{R_{\rm IV}},
\qquad
p \equiv \frac{\bar P_{\rm IV}}{P_0},
\qquad
b \equiv \frac{\rho_0 v_{\rm IV}^2}{P_0},
\end{equation}
Equation~(\ref{eq:stageIV_Rstall_exact}) reduces to:
\begin{equation}\label{eq:stageIV_xstall_poly}
x^{6} - 6p\,x + \left(6p - 1 - b\right) = 0 .
\end{equation}
Using $P_0=\rho_0 c_{s,0}^2/\gamma$ and the Mach number definition (Eq.~\ref{eq:mach_def}), one may write:
\begin{equation}
b = \gamma\,\mathcal{M}_{\rm IV}^2 .
\end{equation}

\paragraph{Small-overshoot approximation.} In the post-equilibrium, subsonic regime characteristic of Stage~IV, it is natural to consider $p=1+\delta$ with $|\delta|\ll1$ and $\mathcal{M}_{\rm IV}\lesssim1$. Writing $x=1+\epsilon$ and assuming $\epsilon\ll1$, we can expand Eq.~\ref{eq:stageIV_xstall_poly} to second order:
\begin{equation}\label{eq:stageIV_eps_quad}
15\,\epsilon^2 - 6\,\delta\,\epsilon - \gamma\,\mathcal{M}_{\rm IV}^2 \simeq 0 .
\end{equation}
The physically relevant root yields:
\begin{equation}\label{eq:stageIV_Rstall_approx}
\begin{aligned}
& \epsilon \simeq \frac{\delta + \sqrt{\delta^2 + \tfrac{5\gamma}{3}\mathcal{M}_{\rm IV}^2}}{5}, \\
& R_{\rm stall} \simeq R_{\rm IV}\,(1+\epsilon).
\end{aligned}
\end{equation}
For near-exact pressure balance at Stage~IV entry ($\delta=0$), a monatomic plasma with $\gamma=5/3$, and a fiducial near-transonic entry $\mathcal{M}_{\rm IV}\simeq 1$, Eq.~\ref{eq:stageIV_Rstall_approx} reduces to:
\begin{equation}\label{eq:stageIV_Rstall_numeric}
\begin{aligned}
R_{\rm stall} & {} \simeq R_{\rm IV}\left[
1 + \sqrt{\frac{\gamma}{15}}\;\mathcal{M}_{\rm IV} \right], \\
& \simeq \frac{4}{3}\,R_{\rm IV}.
\end{aligned}
\end{equation}
Therefore, for $\gamma=5/3$ and a near-transonic Stage~IV entry ($\mathcal{M}_{\rm IV}\simeq 1$) at near-exact pressure balance ($\delta=0$), the bubble could further overshoot by a fractional amount of $\simeq 1/3$.

\paragraph{Approximate stall time.} The stall time follows from the quadrature:
\begin{equation}\label{eq:stageIV_tstall_def}
t_{\rm stall}-t_{\rm IV} = \int_{R_{\rm IV}}^{R_{\rm stall}} \frac{dR}{v(R)},
\end{equation}
where $v(R)$ is given by the Stage~IIIa velocity--radius mapping (Eqs.~\ref{eq:IIIa_first_integral}--\ref{eq:IIIa_v_of_R}) evaluated with $\ast\rightarrow{\rm IV}$.

In the near-equilibrium limit, defining $u\equiv (R-R_{\rm IV})/R_{\rm IV}$, the local expansion of $v(R)$ gives:
\begin{equation}\label{eq:stageIV_v_local}
v(R) \simeq c_{s,0}
\left[\mathcal{M}_{\rm IV}^2 + \frac{6}{\gamma}\,\delta\,u - \frac{15}{\gamma}\,u^2 \right]^{1/2}.
\end{equation}
Substitution into Eq.~\ref{eq:stageIV_tstall_def} yields:
\begin{equation}\label{eq:stageIV_tstall_approx_int}
t_{\rm stall}-t_{\rm IV} \simeq
\frac{R_{\rm IV}}{c_{s,0}}
\int_{0}^{\epsilon}
\frac{du}{\left[\mathcal{M}_{\rm IV}^2 + \frac{6}{\gamma}\,\delta\,u - \frac{15}{\gamma}\,u^2 \right]^{1/2}}.
\end{equation}

For $\delta=0$ and a monatomic plasma with $\gamma=5/3$, the integral evaluates to:
\begin{equation}\label{eq:stageIV_tstall_delta0}
\begin{aligned}
t_{\rm stall}-t_{\rm IV} & {} \simeq
\frac{\pi}{2}\, \sqrt{\frac{\gamma}{15}}\, \frac{R_{\rm IV}}{c_{s,0}}, \\
& \simeq 0.52\,\frac{R_{\rm IV}}{c_{s,0}}.
\end{aligned}
\end{equation}
demonstrating that the duration of the Stage~IV overshoot is of order a fixed fraction of the sound-crossing time at $R_{\rm IV}$. Since Stage~IV begins near the transonic/pressure-balance point, an approximate natural timescale is $t_{\rm IV}\sim R_{\rm IV}/v_{\rm IV} = (R_{\rm IV}/c_{s,0})\,\mathcal{M}_{\rm IV}^{-1}$, implying:
\begin{equation}\label{eq:stageIV_tstall_numeric}
t_{\rm stall} = (1 + \frac{\pi}{6}\,\mathcal{M}_{\rm IV})\,t_{\rm IV} \simeq 1.5\,t_{\rm IV} .
\end{equation}

\paragraph{Practical evaluation of the Stage~IV trajectory.} In practice, the Stage~IV evolution is obtained by continuing the adiabatic pressure closure and momentum-based $v(R)$ mapping inherited from Stage~III, evaluated with the Stage~IV initial conditions $(R_{\rm IV},v_{\rm IV},\bar P_{\rm IV})$. The expansion terminates when the residual momentum is exhausted, defined by $v(R_{\rm stall})=0$, with the corresponding time obtained from a simple quadrature.

For a monatomic plasma ($\gamma=5/3$) entering Stage~IV near pressure balance and transonicity, the analytic approximations derived above yield $R_{\rm stall} \simeq 4/3\,R_{\rm IV}$ (Eq.~\ref{eq:stageIV_Rstall_numeric}) and $t_{\rm stall} \simeq 1.5\,t_{\rm IV}$ (Eq.~\ref{eq:stageIV_tstall_numeric}), implying that Stage~IV contributes only a modest spatial and temporal overshoot beyond the pressure-equilibrium/transonic transition. Throughout this stage, the interior density, temperature, and cooling time are evaluated using the same mass-loading prescription as in Stage~III; for the hot, low-density CGM conditions emphasized here, radiative losses remain dynamically negligible, and the bubble expansion terminates through momentum exhaustion rather than radiative cooling.

%===============================================================
\subsection{Numerical trajectories and comparison of evolutionary branches}
\label{subsec:numerical_trajectories}
%===============================================================

\begin{figure}
\centering
\includegraphics[width=\linewidth]{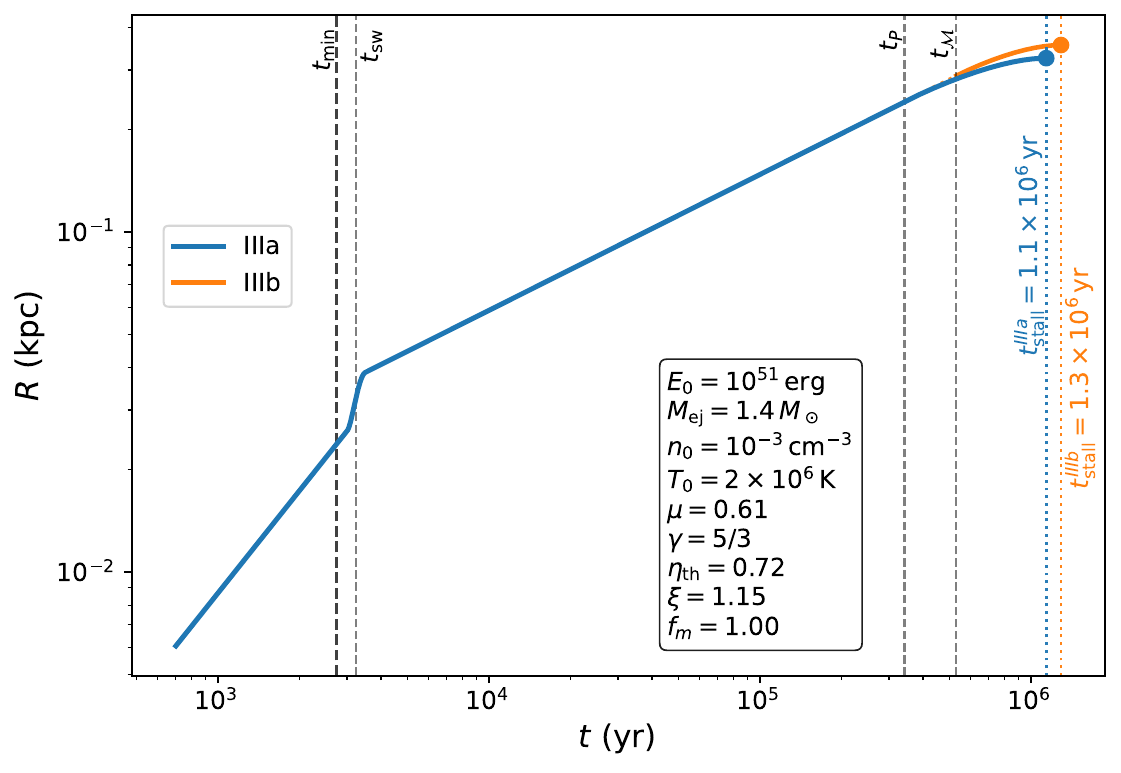}
\caption{Time evolution of the shock radius $R(t)$ for a bubble expanding into a hot CGM. Stages~I (free expansion) and II (Sedov--Taylor) are evaluated analytically and are smoothly joined near the swept-mass equality time $t_{\rm sw}$ (Eq.~\ref{eq:tsw}) using the plotting blend described in the text. We also mark $t_{\min}$ (the leftmost vertical dashed line), defined by the condition $\langle T\rangle_M(t_{\min})=T_{\rm ej}$ (Eqs.~\ref{eq:Tej_bound} and \ref{eq:Tmass_stageI}), and indicates the earliest time at which the Stage~I mean-temperature scaling is physically meaningful. At $t<t_{\min}$, the partition between bulk kinetic and internal energy is still being established, and the plotted curves should be regarded as formal extrapolations rather than literal predictions. Two alternative Stage~III continuations are shown: Stage~IIIa (blue solid), which begins at the pressure-equilibrium time $t_P$, and Stage~IIIb (orange), which begins at the Mach-unity time $t_{\mathcal{M}}$. For clarity, Stage~IIIb is plotted as a dotted curve for $t<t_{\mathcal{M}}$ (formal mathematical continuation) and as a solid curve for $t\ge t_{\mathcal{M}}$ (its physical domain). But the dotted curve is blocked in this plot while visible in a few other plots below. For the fiducial parameters adopted here, $t_P<t_{\mathcal{M}}$, so Stage~IIIa represents the physically realized evolution, while Stage~IIIb is shown for comparison. The stall radius $R_{\rm stall}$ and corresponding time $t_{\rm stall}$ are marked for both branches.}\label{fig:R}
\end{figure}

\begin{figure}
\centering
\includegraphics[width=\linewidth]{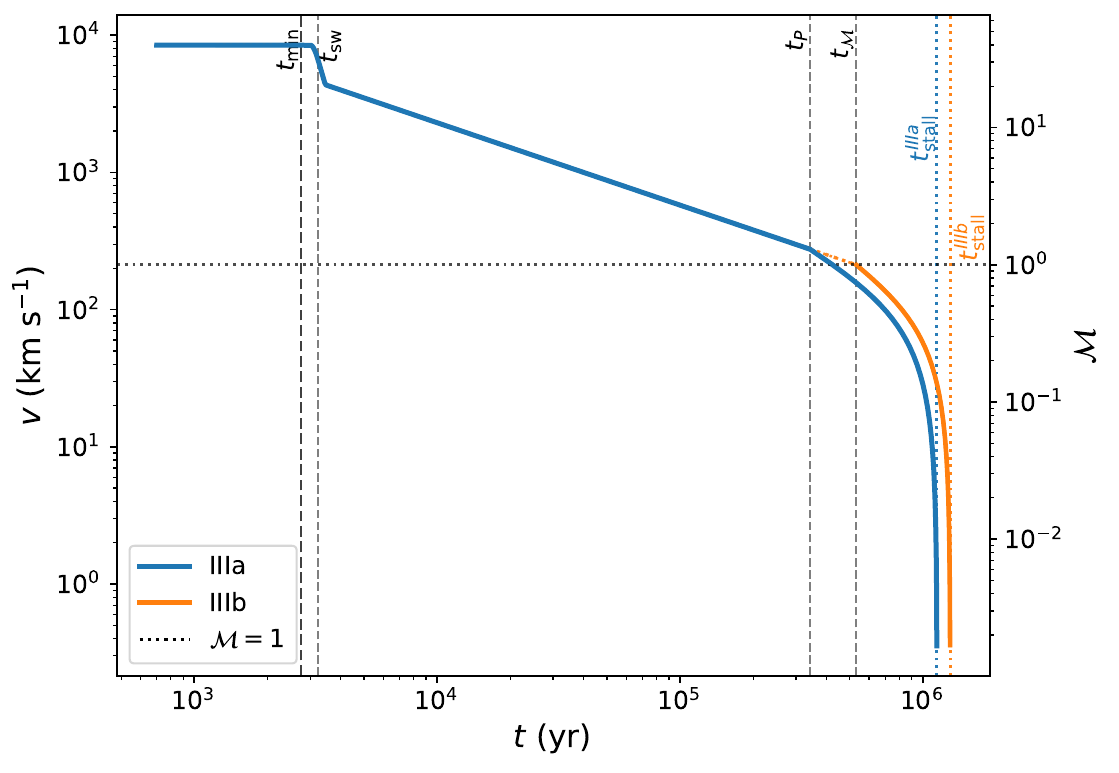}
\caption{Time evolution of the expansion velocity $v(t)$ (left axis) and Mach number $\mathcal{M}(t)=v/c_{s,0}$ (right axis). The horizontal dotted line marks $\mathcal{M}=1$, indicating the transonic transition. Line styles and colors follow the conventions in Fig.~\ref{fig:R}.}\label{fig:vMach}
\end{figure}

\begin{figure}
\centering
\includegraphics[width=\linewidth]{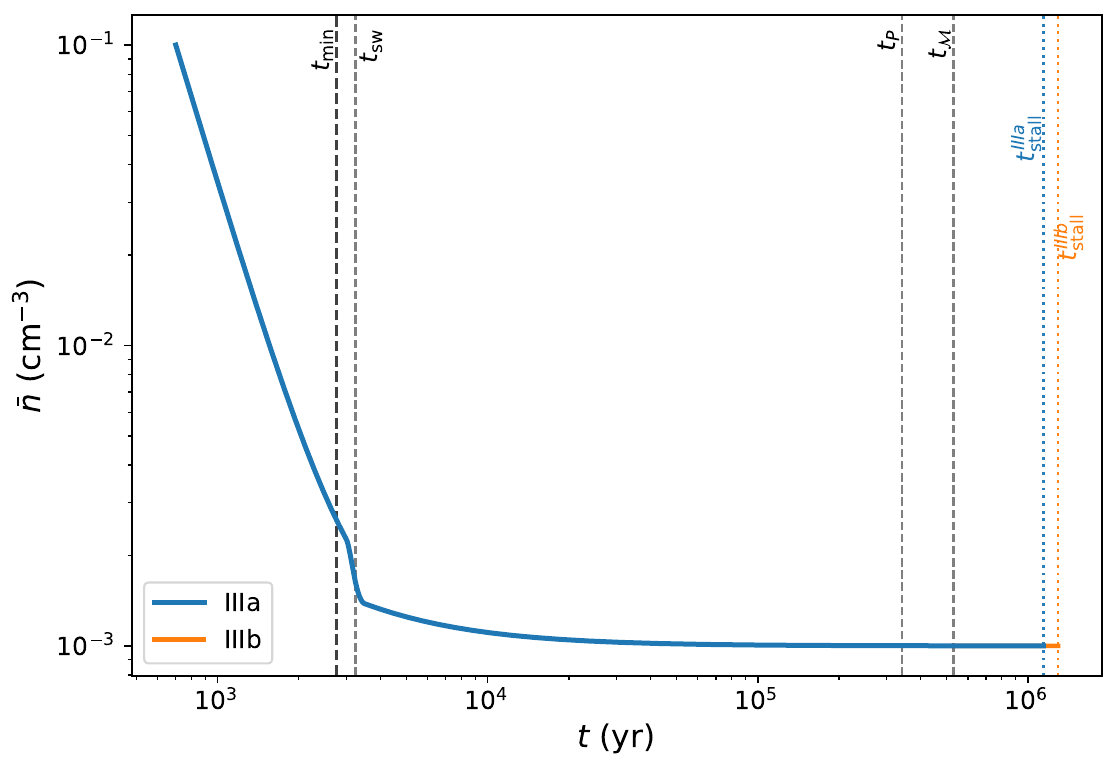}
\caption{Time evolution of the mean interior number density $\bar n(t)$ of the bubble. The calculation includes both ejecta mass and swept-up ambient mass, with a fiducial mass-loading factor $f_m=1$ appropriate for a hot, well-mixed CGM. Line styles and colors follow Fig.~\ref{fig:R}.}\label{fig:nbar}
\end{figure}

\begin{figure}
\centering
\includegraphics[width=\linewidth]{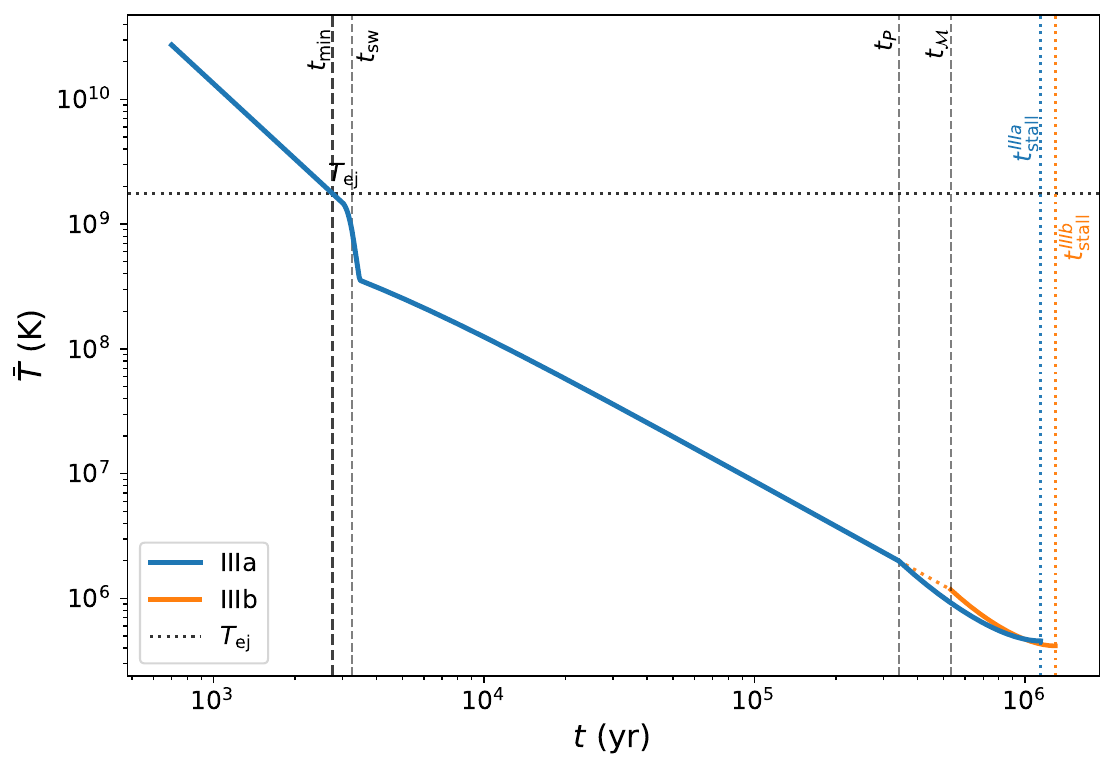}
\caption{Time evolution of the mean interior temperature $\bar T(t)$ of the bubble. The horizontal dotted line indicates the ejecta characteristic temperature $T_{\rm ej}$ (Eq.~\ref{eq:Tej_bound}). The leftmost vertical line marks $t_{\min}$ defined by $\bar T(t_{\min})=T_{\rm ej}$, which we use as an early-time interpretation bound for the Stage~I mean-temperature curve. At $t<t_{\min}$, the plotted increase of $\bar T(t)$ reflects a formal extrapolation of the adopted mean-temperature definition/scaling to arbitrarily early times, when a coherent shocked region and a well-defined thermalized bubble interior have not yet formed; thus $\bar T(t)$ in this regime should not be interpreted as a literal gas temperature. The other line styles and colors follow Fig.~\ref{fig:R}.}\label{fig:Tbar}
\end{figure}

\begin{figure}
\centering
\includegraphics[width=\linewidth]{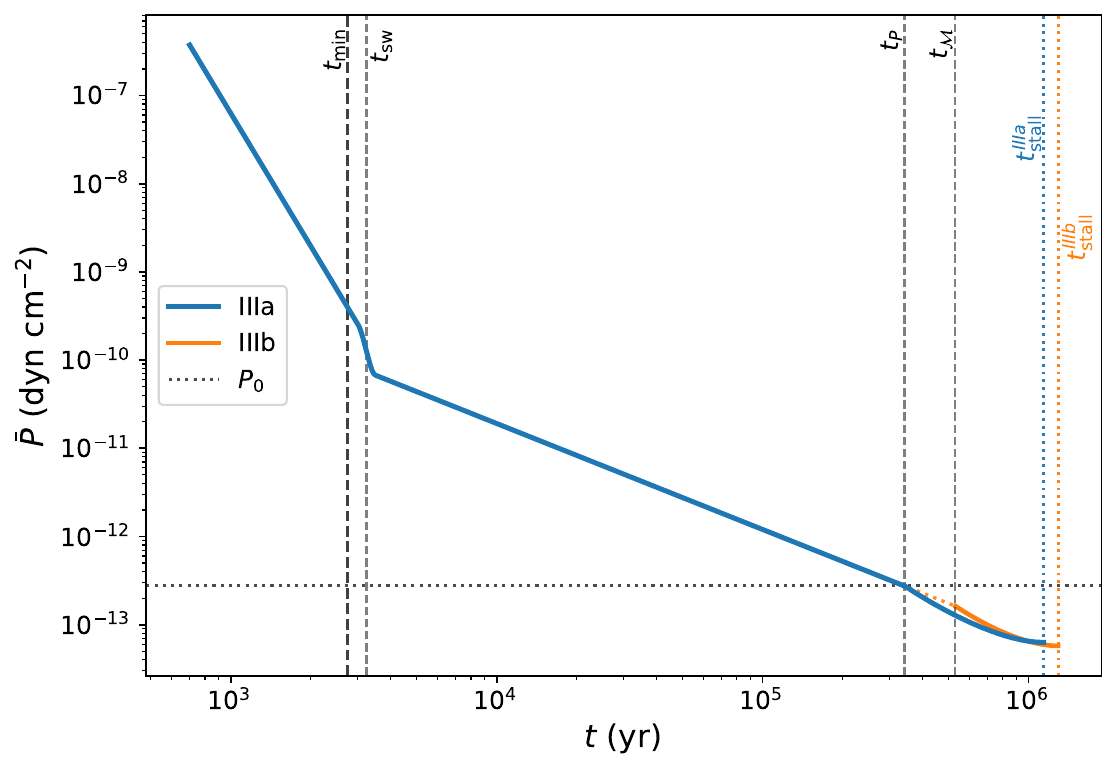}
\caption{Time evolution of the mean interior pressure $\bar P(t)$. The horizontal dotted line marks the ambient thermal pressure $P_0$ of the CGM. The departure from the Sedov--Taylor scaling and the subsequent approach toward pressure equilibrium during Stages~III and IV are clearly visible. Line styles and colors follow Fig.~\ref{fig:R}.}\label{fig:Pbar}
\end{figure}

\begin{figure}
\centering
\includegraphics[width=\linewidth]{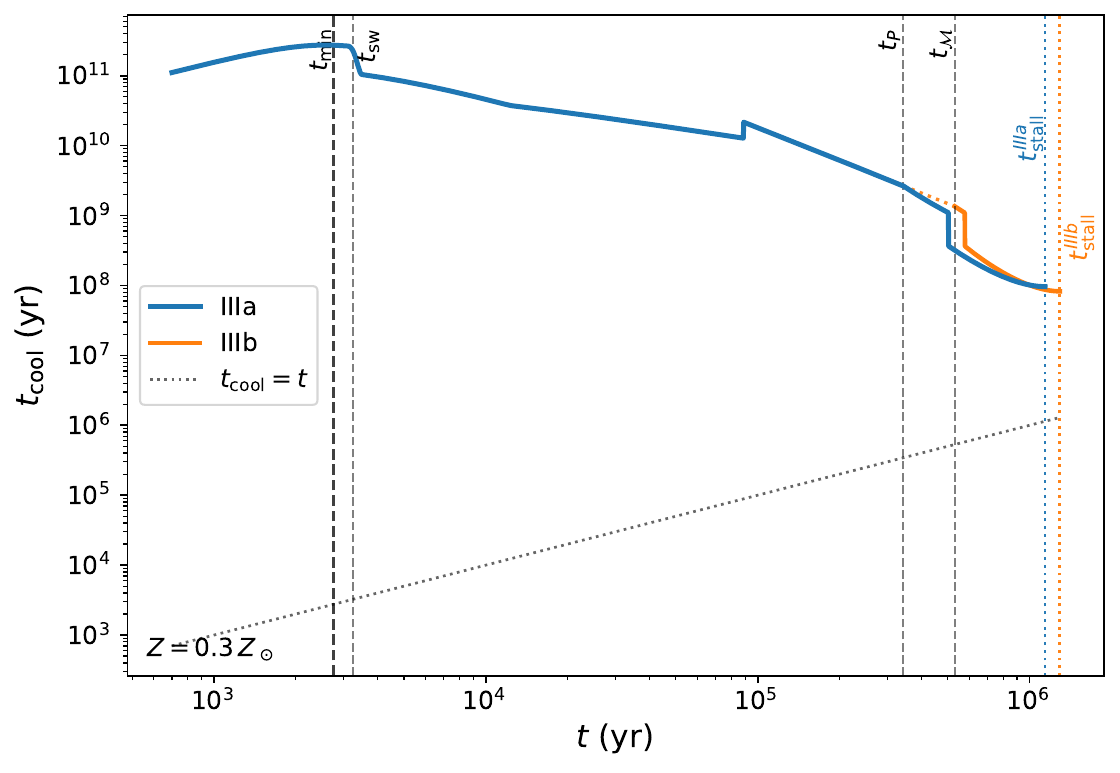}
\caption{Time evolution of the characteristic radiative cooling timescale $t_{\rm cool}(t)$ of the bubble interior, computed from the mean interior density and temperature as $t_{\rm cool}=\tfrac{3}{2}k_B\bar T/[\bar n\,\Lambda(\bar T)]$. For the fiducial hot-CGM parameters adopted here, the cooling time remains much longer than the dynamical time throughout Stages~I--IV; the dotted line marks the locus where the two timescales would be equal. This confirms that radiative losses are dynamically negligible over the evolution shown. Small kinks at late times arise from the temperature dependence of the adopted optically thin cooling function, which is implemented using piecewise power-law approximations based on \citet{Sutherland1993}, rather than from transitions between dynamical stages. Line styles and colors follow Fig.~\ref{fig:R}.}\label{fig:tcool}
\end{figure}

\begin{figure}
\centering
\includegraphics[width=\linewidth]{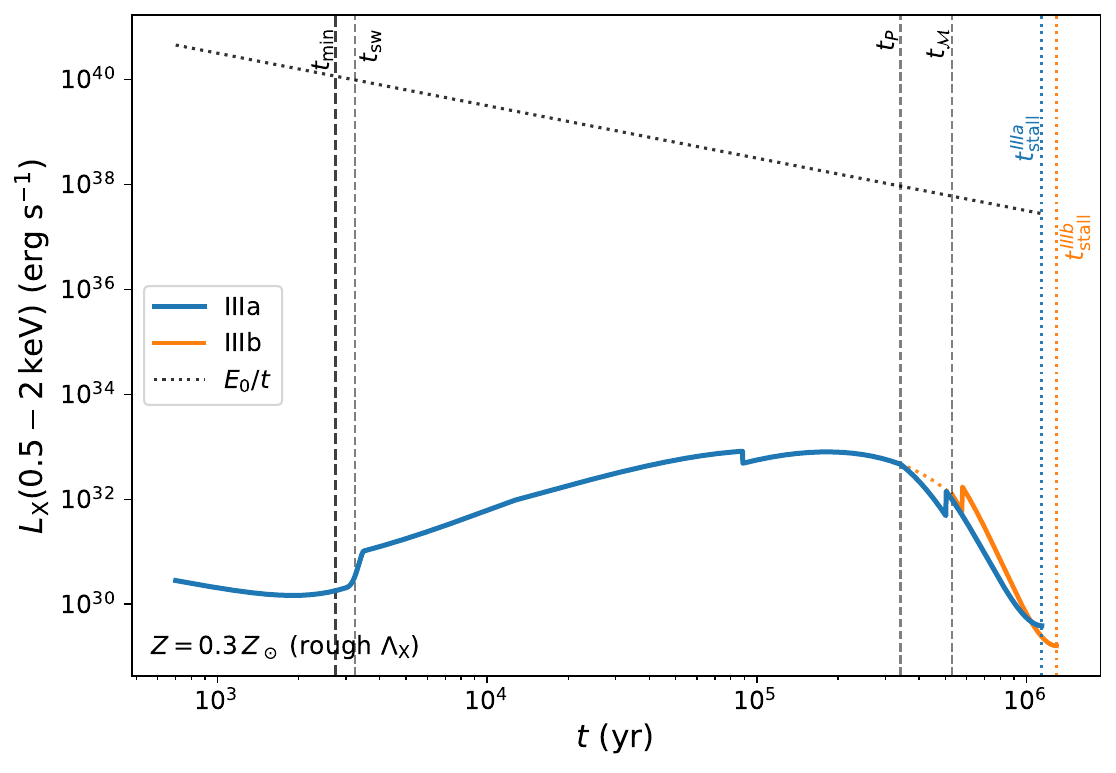}
\caption{Time evolution of the estimated $0.5$--$2$~keV X-ray luminosity $L_X(t)$ of the bubble interior, computed from the mean interior density and temperature as $L_X \simeq \bar n^2 \Lambda_X(\bar T,Z)\,V$, where $V=(4\pi/3)R^3$. We adopt a fiducial metallicity $Z=0.3\,Z_\odot$ and approximate the band-limited emissivity $\Lambda_X$ using an optically thin thermal cooling function restricted to the $0.5$--$2$~keV band. The dashed curve shows $E_0/t$ as a reference scale corresponding to radiating the explosion energy uniformly over time $t$. The large separation between this reference curve and $L_X(t)$ indicates that the bubble is strongly radiatively inefficient, consistent with the long cooling times shown in Fig.~\ref{fig:tcool}. Because the X-ray emissivity scales approximately as $n^2$ while the bubble interior is not spatially uniform, this estimate should be interpreted as an order-of-magnitude diagnostic rather than a precise luminosity prediction. Line styles and colors follow Fig.~\ref{fig:R}.}\label{fig:Lx}
\end{figure}

To provide a unified and quantitative visualization of bubble evolution in a hot CGM, we numerically evaluate and stitch together the time evolution of several key physical quantities, including the shock radius $R(t)$, expansion velocity $v(t)$ (and Mach number $\mathcal{M}\equiv v/c_{s,0}$), mean interior density $\bar n(t)$, temperature $\bar T(t)$, thermal pressure $\bar P(t)$, as well as the cooling time $t_{\rm cool}(t)$, and an order-of-magnitude X-ray luminosity $L_X(t)$. The resulting trajectories, shown in Figs.~\ref{fig:R}--\ref{fig:Lx}, summarize the behavior across Stages~I--IV and enable a direct comparison between the two possible Stage~III continuations discussed in \S\ref{subsec:stageIII_PModifyTransonic}.

\paragraph{Parameter choices and physical normalization.}
Unless otherwise stated, we adopt fiducial parameters representative of a single SN-like energy injection into a hot, low-density CGM: explosion energy $E_0=10^{51}\,{\rm erg}$, ejecta mass $M_{\rm ej}=1.4\,M_\odot$, ambient number density $n_0=10^{-3}\,{\rm cm^{-3}}$, and temperature $T_0=2\times10^6\,{\rm K}$. We assume an ideal monatomic gas with adiabatic index $\gamma=5/3$ and mean molecular weight $\mu=0.61$.

Several dimensionless parameters require further specification. First, the Sedov--Taylor similarity constant $\xi$ is fixed to its standard value for $\gamma=5/3$, $\xi\simeq1.15$, ensuring consistency with the exact similarity solution. Second, we parameterize the relation between the volume-averaged interior pressure and the interior thermal energy during adiabatic evolution using Eq.~\ref{eq:ST_Pbar}. The dimensionless coefficient $\eta_{\rm th}$ accounts for the stratified pressure structure of the Sedov--Taylor solution: the volume-averaged interior pressure is lower than the immediate post-shock pressure by a constant factor. Integrating the exact similarity profiles for $\gamma=5/3$ yields $\eta_{\rm th}$ of order unity, typically in the range $\sim0.6$--$0.8$ (e.g., \citealt{Sedov1959,Chevalier1974,Truelove1999}). We adopt a fiducial value $\eta_{\rm th}=0.72$, while treating it as an adjustable parameter in the numerical evaluation.

We caution that, although the same symbol $\eta_{\rm th}$ is also adopted in Stage~I, its physical meaning differs from that in the Stage~II Sedov--Taylor phase. In Stage~II, self-similarity ensures a fixed partition between thermal and kinetic energy, so that the interior thermal energy remains constant and can be written as $E_{\rm th}=\eta_{\rm th}E_0$ (Eq.~\ref{eq:ST_Pbar}). In contrast, Stage~I is not self-similar, and the interior thermal energy is an explicitly time-dependent, Eulerian quantity that declines with expansion as $E_{\rm th}\propto R^{-2}$ (Eq.~\ref{eq:EthED}). In this case, $\eta_{\rm th}$ should be interpreted only as a normalization inherited from the Sedov--Taylor solution, while the time variation of the effective thermal fraction is fully captured by Eq.~\ref{eq:EthED}.

When computing the mean interior density, we include a mass-loading parameter $f_m$ such that the total hot gas mass inside the bubble is $M_{\rm tot}=M_{\rm ej}+f_m M_{\rm sw}$, where $M_{\rm sw}$ is the swept-up ambient mass (Eq.~\ref{eq:Mb_fm_norm}). In a hot CGM, where the shocked ambient gas can efficiently mix with the bubble interior, we adopt $f_m=1$ as a representative choice.

\paragraph{Early-time validity bound for Stage~I mean temperatures.} The Stage~I scaling $E_{\rm th}\propto R^{-2}$ (Eq.~\ref{eq:EthED}) and the corresponding mean temperatures (e.g., Eq.~\ref{eq:Tmass_stageI}) should be regarded as intermediate-asymptotic relations that apply once a coherent shocked region and a well-defined bubble interior have formed. They are not intended to be extrapolated to arbitrarily early times ($t\rightarrow 0$), when the partition between bulk kinetic energy and internal energy is still being established and when additional microphysical processes -- such as incomplete thermalization, electron-ion non-equilibration, or a finite ejecta thickness or launch scale -- may be important.

A practical lower bound on the applicability of the Stage~I temperature scaling can be obtained by comparing the mean interior temperature to the characteristic specific energy of the ejecta. Defining an ejecta characteristic velocity and temperature using Eqs.~\ref{eq:vej_bound} and \ref{eq:Tej_bound}, we introduce a minimum time $t_{\min}$ by the condition $\langle T\rangle_M(t_{\min}) = T_{\rm ej}$. For the fiducial model adopted in this paper, we find $T_{\rm ej}\simeq1.8\times10^9\,{\rm K}$ and $t_{\min}\simeq2.7\times10^3\,{\rm yr}$.

In all figures, we mark $t_{\min}$ explicitly and restrict the displayed time range to $t\simeq7\times10^2\,{\rm yr}$--$2\times10^6\,{\rm yr}$. Results at $t<t_{\min}$ are shown only for completeness and should not be interpreted literally. Since the early evolutionary phases (Stages~I and II) closely resemble the standard SNR evolution (e.g., \citealt{Sedov1959,Truelove1999}), our analysis focuses on the subsequent stages at times well beyond $t_{\min}$. Consequently, the unphysical thermal state implied by the Stage~I scaling at $t<t_{\min}$ does not affect the physical interpretation or conclusions presented below.

\paragraph{Treatment of Stages~I and II and continuity.}
Stages~I (free expansion) and II (Sedov--Taylor) admit closed-form analytical solutions and are evaluated directly from the corresponding expressions presented in \S\ref{subsec:StageI_ED} and \S\ref{subsec:stageII_ST}. In Stage~I, the shock radius and velocity are given by Eqs.~\ref{eq:StageIRtvt}. The interior thermal state is characterized using representative mean quantities derived from the explicitly time-dependent thermal energy $E_{\rm th}(t)$ (Eq.~\ref{eq:EthED}), which is defined as the instantaneous total thermal energy contained within the bubble interior. The mean interior density is computed from the total gas mass (Eq.~\ref{eq:MtotMsw}) and the bubble volume, while the mean pressure is evaluated consistently via the relation $\bar P=(\gamma-1)E_{\rm th}/V$. The corresponding Stage~I mean temperature is then obtained using the mass-weighted definition in Eq.~\ref{eq:Tmass_stageI}.

In Stage~II, the evolution of the shock radius and velocity follows the Sedov--Taylor similarity solution (Eq.~\ref{eq:ST_R_v}). Owing to self-similarity, the partition of the explosion energy between thermal and kinetic components becomes time-invariant, and the interior thermal energy approaches a constant fraction of the total explosion energy. The mean interior pressure and temperature are therefore evaluated using the Sedov--Taylor relations in Eqs.~\ref{eq:ST_Pbar} and \ref{eq:ST_Tbar_general}, respectively.

The transition between Stages~I and II occurs near the swept-mass equality time $t_{\rm sw}$ defined in Eq.~\ref{eq:tsw}. When the Stage~I and Stage~II solutions are evaluated independently, they do not in general match exactly at $t_{\rm sw}$, which can introduce small but artificial discontinuities or cusps in $R(t)$ and especially in $v(t)$ when plotted. To avoid such visual artifacts, we apply a narrow logarithmic blending in time centered on $t_{\rm sw}$ when constructing the figures. Specifically, over a small interval symmetric in $\log t$ around $t_{\rm sw}$, we smoothly interpolate between the Stage~I and Stage~II solutions for plotting purposes only, ensuring continuity of both $R(t)$ and $v(t)$. Outside this narrow interval, the analytical Stage~I and Stage~II solutions are used exactly as derived. This procedure does not modify the physical scalings, introduce a new dynamical phase, or affect the subsequent evolution; it serves solely to improve the clarity of the numerical visualization.

\paragraph{Numerical evaluation of Stages III and IV.}
Beyond the Sedov--Taylor phase, the evolution is governed by the pressure-modified dynamics described in \S\ref{subsec:stageIII_PModifyTransonic}. 
For Stage~III, we evaluate the trajectory numerically using the first-integral mapping between the expansion velocity $v$ and radius $R$ derived from the equation of motion including ambient pressure effects (Eq.~\ref{eq:momentum}), which yields the velocity--radius relation given in Eq.~\ref{eq:IIIa_v_of_R}. The interior pressure is evolved using the adiabatic closure $\bar P(R)\propto R^{-3\gamma}$ as given in Eq.~\ref{eq:IIIa_PofR_general}. The corresponding time coordinate is obtained by numerically integrating $dt=dR/v(R)$, as expressed explicitly in Eq.~\ref{eq:IIIa_t_of_R}. This procedure remains valid until the expansion velocity formally vanishes at a finite radius.

Two alternative Stage~III branches are computed and overplotted. Stage~IIIa begins at the pressure-equilibrium time $t_P$, defined by Eqs.~\ref{eq:Pbttimescale} and \ref{eq:tPnorm}, and follows the pressure-modified evolution described above. Stage~IIIb begins at the Mach-unity time $t_{\mathcal{M}}$, defined by Eqs.~\ref{eq:tM} and \ref{eq:tMnorm}, and is governed by the same equation of motion but initialized with the Mach-unity boundary condition. For completeness, we also evaluate Stage~IIIb independently using the equivalent ordinary differential equation for $v(R)$ given in Eq.~\ref{eq:IIIb_ODE}, which yields identical trajectories within numerical precision. For the fiducial parameter set shown here, $t_P<t_{\mathcal{M}}$, so Stage~IIIa represents the physically realized continuation, while Stage~IIIb is shown only for $t\ge t_{\mathcal{M}}$ as a comparison case.

Stage~IV corresponds to the subsequent subsonic relaxation toward pressure equilibrium with the ambient CGM. In this regime, the expansion velocity decreases rapidly below the ambient sound speed and the bubble ceases to expand significantly. In practice, the numerical integration naturally terminates at a stall radius $R_{\rm stall}$ and corresponding time $t_{\rm stall}$, defined by the condition $v=0$ in Eq.~\ref{eq:momentum} and evaluated explicitly using Eqs.~\ref{eq:Rstall_def} and \ref{eq:tstall_def}. These quantities are marked explicitly for both Stage~III branches in the figures. Beyond this point, the bubble effectively merges into the background CGM and no longer constitutes a dynamically distinct structure.

\paragraph{Plotting convention for the two Stage~III branches.}
For consistency across all diagnostic figures, we show both Stage~III continuations in every panel. The branch that is physically realized is determined by the ordering of $t_P$ and $t_{\mathcal{M}}$. The alternative branch is plotted as a formal continuation initialized at its own transition time. In particular, Stage~IIIb is only physically applicable for $t\ge t_{\mathcal{M}}$; nevertheless, its governing equations admit a mathematical continuation to earlier times. We therefore plot Stage~IIIb as a dotted curve for $t<t_{\mathcal{M}}$ and as a solid curve for $t\ge t_{\mathcal{M}}$. This convention is applied uniformly to $R(t)$, $v(t)$ (and $\mathcal{M}$), $\bar n(t)$, $\bar T(t)$, $\bar P(t)$, and later also to $t_{\rm cool}(t)$ and $L_X(t)$.

\paragraph{Radiative cooling timescale.}
In addition to the dynamical quantities, we compute the characteristic radiative cooling timescale of the hot bubble interior following Eq.~\ref{eq:tcool}, and adopt the bubble interior average parameters $\bar T(t)$ and $\bar n(t)$, as well as a standard collisional-ionization-equilibrium cooling curve appropriate for CGM conditions (e.g., \citealt{Sutherland1993}) evaluated at a fiducial metallicity $Z=0.3Z_\odot$.

Fig.~\ref{fig:tcool} shows $t_{\rm cool}(t)$ for both Stage~III branches. Throughout Stages~I--III, the cooling time remains orders of magnitude longer than the dynamical time $t$, justifying the adiabatic treatment adopted in our modeling. Only at late times, as the bubble approaches pressure equilibrium and the mean interior temperature declines, does $t_{\rm cool}$ begin to decrease appreciably, though radiative losses remain dynamically subdominant for the parameters considered here.

Several small kinks visible in the $t_{\rm cool}(t)$ curve are not associated with dynamical transitions. Instead, they arise from the temperature dependence of the adopted cooling function, which is implemented using piecewise power-law approximations across different temperature ranges \citep{Sutherland1993}. As $\bar T(t)$ crosses the characteristic temperatures separating these regimes, the cooling function changes slope, producing corresponding features in $t_{\rm cool}(t)$ despite the smooth evolution of $\bar n(t)$ and $\bar T(t)$.

\paragraph{Order-of-magnitude X-ray luminosity estimate.}
Using the numerically evaluated mean interior density and temperature, we derive an approximate $0.5$--$2$~keV X-ray luminosity,
\begin{equation}
L_X(t) \;\simeq\; \bar n^2(t)\,\Lambda_X[\bar T(t),Z]\,V(t),
\end{equation}
where $V=(4\pi/3)R^3$ and $\Lambda_X(T,Z)$ denotes a band-limited optically thin emissivity. In practice, $\Lambda_X$ is approximated by restricting the cooling function to the $0.5$--$2$~keV band for a fiducial metallicity $Z=0.3\,Z_\odot$. We also overplot $E_0/t$ as a reference scale corresponding to radiating the explosion energy over time $t$.

This $L_X$ estimate is intended only as an order-of-magnitude diagnostic. Because the interior density is not spatially uniform and the emissivity scales approximately as $n^2$, using $\bar n^2 V$ generally underestimates contributions from denser regions (e.g., near the shell or in compressed substructures) and overestimates those from tenuous volume-filling gas. A more accurate prediction would require the full density and temperature distributions, which are beyond the scope of the present analytic framework. Nevertheless, the comparison between $L_X(t)$ and $E_0/t$ provides a useful gauge of the overall X-ray radiative efficiency during the different evolutionary stages. As shown in Fig.~\ref{fig:Lx}, even the peak $L_X$ accounts for $\lesssim0.1\%$ of the explosion energy release rate. Considering the density structure of the bubble interior, this fraction may be higher, but still orders of magnitude lower than the explosion energy. This ensures the inefficiency of X-ray radiation in a hot CGM environment, qualitatively consistent with numerical simulations \citep{Tang2005,Tang2009} and X-ray observations (e.g., \citealt{LiJ2011,LiJ2017}).

%===============================================================
\section{Global observable outcomes}
\label{sec:global_outcomes}
%===============================================================

In \S\ref{sec:baselinemodel} we developed a fully time-dependent baseline model for the evolution of a single, instantaneous energy injection expanding into a uniform hot CGM. While that analysis is essential for understanding the detailed dynamics, many observational and theoretical applications are better framed in terms of a small number of \emph{global} or \emph{integrated} quantities, such as the maximum spatial extent, effective lifetime, total radiative output, and characteristic ionic column densities.

In this section we therefore compress the evolutionary trajectories discussed in \S2 into a set of physically motivated global observables. Our goal is to identify how the bubble properties are regulated by the input parameters, regardless of the uncertain microphysics. For simplicity, we adopt a fiducial CGM temperature of $T_0 = 0.5~{\rm keV}$, and explore the dependence on the explosion energy $E_0$ and ambient number density $n_0$. At fixed $T_0$, the ambient pressure $P_0 = n_0 k_B T_0$ is the dominant thermodynamic control parameter, so variations in $n_0$ capture most of the relevant environmental dependence for hot halos.

%---------------------------------------------------------------
\subsection{Maximum radius and effective lifetime}
\label{subsec:Rmax_lifetime}
%---------------------------------------------------------------

A central question for feedback in a hot CGM is how far the influence of a single energetic event extends, and for how long it persists. In a pressurized, high--sound-speed medium, this question admits two distinct but complementary answers, depending on whether one is concerned with the \emph{detectability} of a coherent shell or the \emph{ultimate deposition of energy} into the ambient medium.

\subsubsection{Detectable-shell radius and lifetime}\label{subsubsec:Rdet_tdet}

We first define a \emph{detectable-shell radius}, $R_{\rm det}$, as the maximum radius at which the bubble can be identified as a coherent structure with a well-defined shell or cavity. In the evolutionary framework of \S\ref{sec:baselinemodel}, this corresponds to the termination of Stage~III, when either pressure equilibrium with the ambient CGM is reached or the expansion becomes transonic:
\begin{equation}\label{eq:Rdettdet}
R_{\rm det} \equiv R(t_{\rm det}),
\qquad
t_{\rm det} \equiv \max\!\left(t_P,\, t_{\mathcal{M}}\right),
\end{equation}
where $t_P$ and $t_{\mathcal{M}}$ are defined in Eqs.~\ref{eq:tPnorm} and \ref{eq:tMnorm}, respectively. Beyond $t_{\rm det}$, the forward shock degenerates into a weak compression wave, the pressure contrast across the shell drops to order unity, and the surface brightness of any shell-like structure declines rapidly. As a result, $R_{\rm det}$ sets the characteristic maximum \emph{observable} size of an individual bubble in X-ray emission or other density-sensitive tracers.

For a fixed ambient temperature $T_0$, both $t_P$ and $t_{\mathcal{M}}$ scale primarily with the ratio $E_0/P_0$. To leading order this implies:
\begin{equation}
R_{\rm det} \propto \left(\frac{E_0}{P_0}\right)^{1/3}
\propto E_0^{1/3} n_0^{-1/3},
\end{equation}
up to order-unity factors involving the thermalization efficiency $\eta_{\rm th}$, the Sedov--Taylor constant $\xi$, and the fixed $T_0 = 0.5~{\rm keV}$. This scaling emphasizes that, in a hot CGM, the detectable extent of feedback is regulated by ambient pressure rather than by radiative cooling, in contrast to the classical cold-ISM picture. 

\begin{figure}
  \centering
  \includegraphics[width=\linewidth]{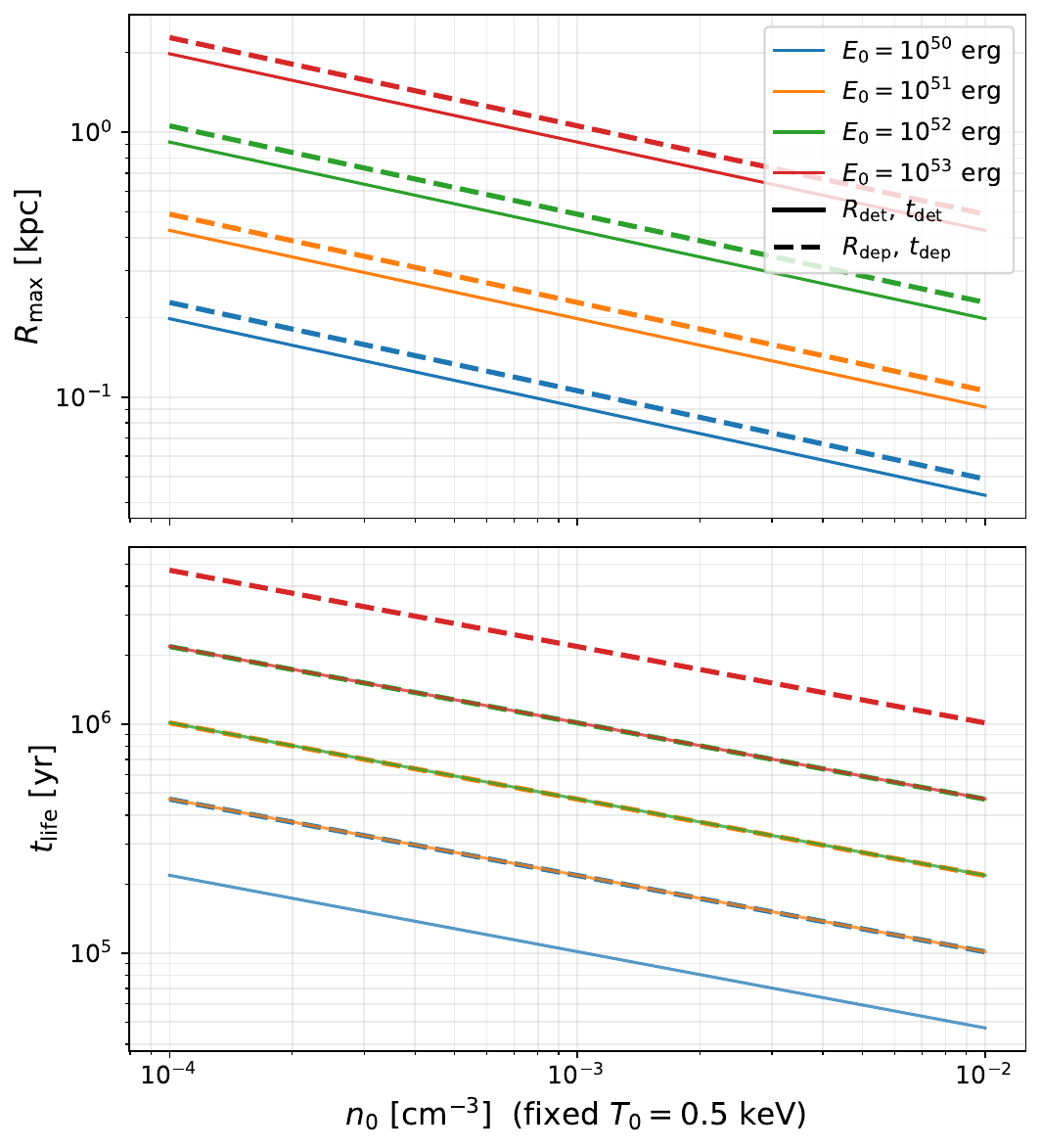}
  \caption{Global spatial reach and effective lifetime of a single-event bubble in a hot CGM as a function of ambient density $n_0$, at fixed $T_0=0.5$~keV. \textit{Top:} Maximum radius $R_{\max}$ in two complementary senses: the \emph{detectable-shell} scale $R_{\rm det}$ (solid), defined at $t_{\rm det}\equiv\max(t_P,t_{\mathcal{M}})$ when the expansion becomes pressure-modified/transonic and a coherent shell rapidly fades, and the \emph{energy-deposition} scale $R_{\rm dep}$ (dashed), taken as the Stage~IV stall radius $R_{\rm stall}$ that characterizes the ultimate volume over which the injected energy is deposited into the CGM. \textit{Bottom:} Corresponding characteristic times, $t_{\rm det}$ (solid) and $t_{\rm dep}\equiv t_{\rm stall}$ (dashed). Colors indicate the explosion energy $E_0$ (as labeled).}\label{fig:global_Rmax_lifetime}
\end{figure}

In practice, $R_{\rm det}$ is evaluated using the same numerical procedure described in \S\ref{subsec:numerical_trajectories}. Specifically, for a given set of parameters, the bubble trajectory $R(t)$ is constructed by stitching the Stage~II solution to the appropriate Stage~III continuation, and the detectable-shell radius is obtained by evaluating $R(t)$ at $t=t_{\rm det}$. This procedure is identical to that used to generate the trajectories shown in Fig.~\ref{fig:R}. The resulting values are summarized in Fig.~\ref{fig:global_Rmax_lifetime} as solid curves, where $R_{\rm max}$ denotes either $R_{\rm det}$ or the deposition radius $R_{\rm dep}$ defined in \S\ref{subsubsec:Rdep_tdep}, and $t_{\rm life}$ denotes the corresponding characteristic time scale, $t_{\rm det}$ or $t_{\rm dep}$. 

We emphasize that the term ``detectable'' is used here in a purely physical and morphological sense. The existence of a well-defined shell or cavity at $R_{\rm det}$ does not guarantee an actual observational detection, which in practice depends on the intrinsic surface brightness of the structure, its contrast relative to the background, and the sensitivity and angular resolution of a given observation. Rather, $R_{\rm det}$ characterizes the maximum radius at which a bubble remains, in principle, distinguishable from the ambient CGM given sufficiently deep observations.

\subsubsection{Energy-deposition radius and lifetime} \label{subsubsec:Rdep_tdep}

The loss of a coherent shell does not mark the end of the dynamical influence of the bubble. After entering Stage~IV, the disturbance continues to redistribute energy through subsonic expansion, weak compression waves, and residual momentum, even though a well-defined shell or cavity may no longer be present.

We therefore define a second characteristic scale, the \emph{energy-deposition radius}, $R_{\rm dep}$, as the maximum radius reached during the post-transonic evolution:
\begin{equation}
R_{\rm dep} \equiv R_{\rm stall},
\qquad
t_{\rm dep} \equiv t_{\rm stall},
\end{equation}
where $R_{\rm stall}$ and $t_{\rm stall}$ are defined in \S\ref{subsec:stageIV_subsonic}. These quantities characterize the point at which the expansion effectively stalls and the remaining energy has been deposited into the surrounding CGM.

As shown in \S\ref{subsec:stageIV_subsonic}, the stall radius and time are well approximated by:
\begin{equation}
R_{\rm dep} \simeq \frac{4}{3}\,R_{\rm IV},
\qquad
t_{\rm dep} \simeq 1.5\,t_{\rm IV},
\end{equation}
indicating that a non-negligible fraction of the spatial extent is achieved after the expansion has become fully subsonic.

Physically, $R_{\rm dep}$ characterizes the volume over which the injected energy is ultimately deposited into the CGM. This scale is therefore more relevant for questions of entropy injection, thermalization, and large-scale CGM heating than $R_{\rm det}$, even though it is not associated with a morphologically distinct shell or cavity.

The quantities $R_{\rm dep}$ and $t_{\rm dep}$ are shown as dashed curves in Fig.~\ref{fig:global_Rmax_lifetime}. We explore the maximum spatial extent and effective lifetime of expanding bubbles/superbubbles over a range of explosion energies, $E_0=10^{50}$--$10^{53}\,{\rm erg}$, and ambient CGM densities, $n_0=10^{-4}$--$10^{-2}\,{\rm cm^{-3}}$. For typical parameters, individual bubbles reach characteristic sizes of order $\sim$kpc and persist for $\sim$Myr.

These lifetimes are comparable to or shorter than those of typical star-forming regions, while the maximum radii are generally smaller than the spatial extent of the diffuse X-ray--emitting CGM detected around nearby galaxies (e.g., \citealt{LiJ2013a,LiJ2017}). This implies that large-scale galactic outflows are more likely driven by collective and continuous energy injection from multiple events rather than by isolated SNe in actively star forming galaxies. In contrast, in environments such as galactic bulges where intense star formation is absent, feedback from individual SNe occurring at different locations may instead play a dominant role, giving rise to buoyant motions and subsonic disturbances (e.g., \citealt{Tang2005,Tang2009,LiJ2009,LiJ2011}).

%---------------------------------------------------------------
\subsection{X-ray luminosity, cooling, and energy deposition}
\label{subsec:Lx_cooling}
%---------------------------------------------------------------

The hot CGM environment fundamentally alters the relationship between radiative output and dynamical impact. Here we reinterpret the X-ray luminosity and cooling-time results from \S\ref{subsec:numerical_trajectories} in terms of global energetic efficiency and feedback deposition.

\subsubsection{Cooling versus dynamics}\label{subsubsec:coolingvsdynamics}

As demonstrated in \S\ref{subsec:numerical_trajectories} and Fig.~\ref{fig:tcool}, the characteristic cooling time of the hot bubble interior generally satisfies $t_{\rm cool} \gg t_{\rm dyn}$ over most of the evolution ($t_{\rm cool}$ before $t_{\rm min}$ in Stage~I is unphysical) for fiducial CGM densities and temperatures. This inequality typically holds through Stages~II--IV unless extreme mass loading or metallicity enhancement is assumed. The implication is that, for a single SN-like event in a hot CGM, radiative losses do not regulate the expansion or termination of the bubble; instead, the evolution is controlled primarily by pressure confinement and the transition to transonic and subsonic flow.

To connect cooling to observability, we evaluate the ratio $t_{\rm cool}/t_{\rm dyn}$ at the ``detectable-shell'' epoch $t_{\rm det}$ and plot it in the upper panel of Fig.~\ref{fig:global_cooling_efficiency}. Here $t_{\rm dyn}\equiv R/v$ is taken from the dynamical trajectory, while $t_{\rm cool}$ is computed from the same optically thin prescription used throughout this work (Eq.~\ref{eq:tcool}), which emphasizes that cooling depends on both the thermodynamic state $(T_b,n_b)$ and the temperature dependence of the cooling function $\Lambda$.

A notable trend in Fig.~\ref{fig:global_cooling_efficiency} is that, at fixed $n_0$ ($T_0$ so $P_0$ are also fixed), increasing the injected energy $E_0$ tends to \emph{reduce} $t_{\rm cool}/t_{\rm dyn}$ evaluated at $t=t_{\rm det}$. This behavior does \emph{not} mean that higher-$E_0$ explosions are intrinsically ``more radiative'' at a fixed evolutionary time; rather, it reflects how the \emph{evaluation epoch} shifts with $E_0$. More energetic events maintain a detectable shell to later times (larger $t_{\rm det}$). Since the flow is decelerating, the characteristic dynamical time at that epoch is longer, with $t_{\rm dyn}\sim R/v$ typically of order the age $t$ up to stage-dependent factors (e.g., $R/v = (5/2)t$ for a pure Sedov--Taylor scaling). Thus, even if $t_{\rm cool}$ varied weakly with $E_0$, one generically expects:
\begin{equation}
\left.\frac{t_{\rm cool}}{t_{\rm dyn}}\right|_{t_{\rm det}}
\propto
\left.\frac{t_{\rm cool}}{t}\right|_{t_{\rm det}}
\end{equation}
to decrease as $t_{\rm det}$ shifts to larger values for larger $E_0$.

In addition, the thermodynamic state at $t_{\rm det}$ also changes systematically with $E_0$. Because the interior temperature declines with time along an adiabatic trajectory, a larger $t_{\rm det}$ implies a lower $T_b(t_{\rm det})$. In the relevant $T\sim 10^{6}$--$10^{7}$~K regime, $\Lambda(T,Z)$ is generally \emph{not} constant: as $T_b$ decreases toward the metal-line cooling peak, $\Lambda$ typically increases, which further shortens $t_{\rm cool}$ (Figs.~\ref{fig:Tbar}, \ref{fig:tcool}; \citealt{Sutherland1993}). These two effects---a longer dynamical time and a modestly enhanced cooling function at later times---act in the same direction, yielding a smaller $t_{\rm cool}/t_{\rm dyn}$ at $t_{\rm det}$ for larger $E_0$.

Astronomically, this trend implies that the most energetic superbubbles, although still largely adiabatic in the hot CGM at the time their shells become hard to detect, are \emph{closer} to the regime where radiative losses and mixing can compete with dynamics near the end of their observable lifetime. In practice, the trend in Fig.~\ref{fig:global_cooling_efficiency} formally suggests that higher-$E_0$ events move closer to a regime where cooling-assisted mixing or condensation could in principle become relevant after the shell fades. However, for the physical parameter range considered here the cooling time of the hot CGM remains orders of magnitude longer than the dynamical time, so such a phase is unlikely to occur in realistic systems unless extreme conditions (e.g., unrealistically large $E_0$ or much higher ambient densities) are invoked. This provides a natural pathway for energetic feedback episodes to leave behind transient, cooling-influenced CGM signatures (e.g., enhanced soft-X-ray emissivity and intermediate-temperature interfaces) without requiring that radiative cooling dominate the expansion history itself. This scenario is especially plausible when significant cold-gas mass loading and enhanced radiative cooling are present, as is often the case for intense energy injection driven by star-forming galaxies, where entrainment and mixing of cooler material can substantially reduce the effective cooling time (e.g., \citealt{Strickland2000,LiJ2009,LiJ2011,LiJ2013b}).

\subsubsection{Radiative inefficiency versus dynamical impact}
\label{subsubsec:RadiativeInefficiencyDynamicalImpact}

We define a global measure of radiative efficiency as:
\begin{equation}\label{eq:RadiativeEfficiencyBubble}
\epsilon_X \equiv \frac{1}{E_0}\int_{t_{\min}}^{t_{\rm stall}} L_X(t)\,dt,
\end{equation}
which quantifies the fraction of the injected energy emitted as X-rays over the physically applicable portion of the evolution. Here $t_{\min}$ is a conservative early-time validity bound for our thermal-state estimates, defined in \S\ref{subsec:numerical_trajectories} by the condition $\langle T\rangle_M(t_{\min})=T_{\rm ej}$, so that for $t<t_{\min}$ the formal Stage~I mean temperature would exceed the temperature scale implied by the ejecta specific energy and should not be regarded as physical. The upper limit $t_{\rm stall}$ denotes the maximum-radius (zero-velocity) epoch at which the outward expansion halts ($v=0$ at $R_{\rm stall}$; see \S\ref{subsec:numerical_trajectories}). Note that $t_{\rm stall}$ marks the end of \emph{outward expansion}; residual energy can still be redistributed by subsonic pressure equilibration and wave-like disturbances even after $R(t)$ ceases to increase. Here $L_X$ denotes the $0.5$--$2$~keV band luminosity estimated from $\bar n^2 \Lambda_X(\bar T,Z)V$ using the same band-limited emissivity model as in \S\ref{subsec:numerical_trajectories}.

\begin{figure}
  \centering
  \includegraphics[width=\linewidth]{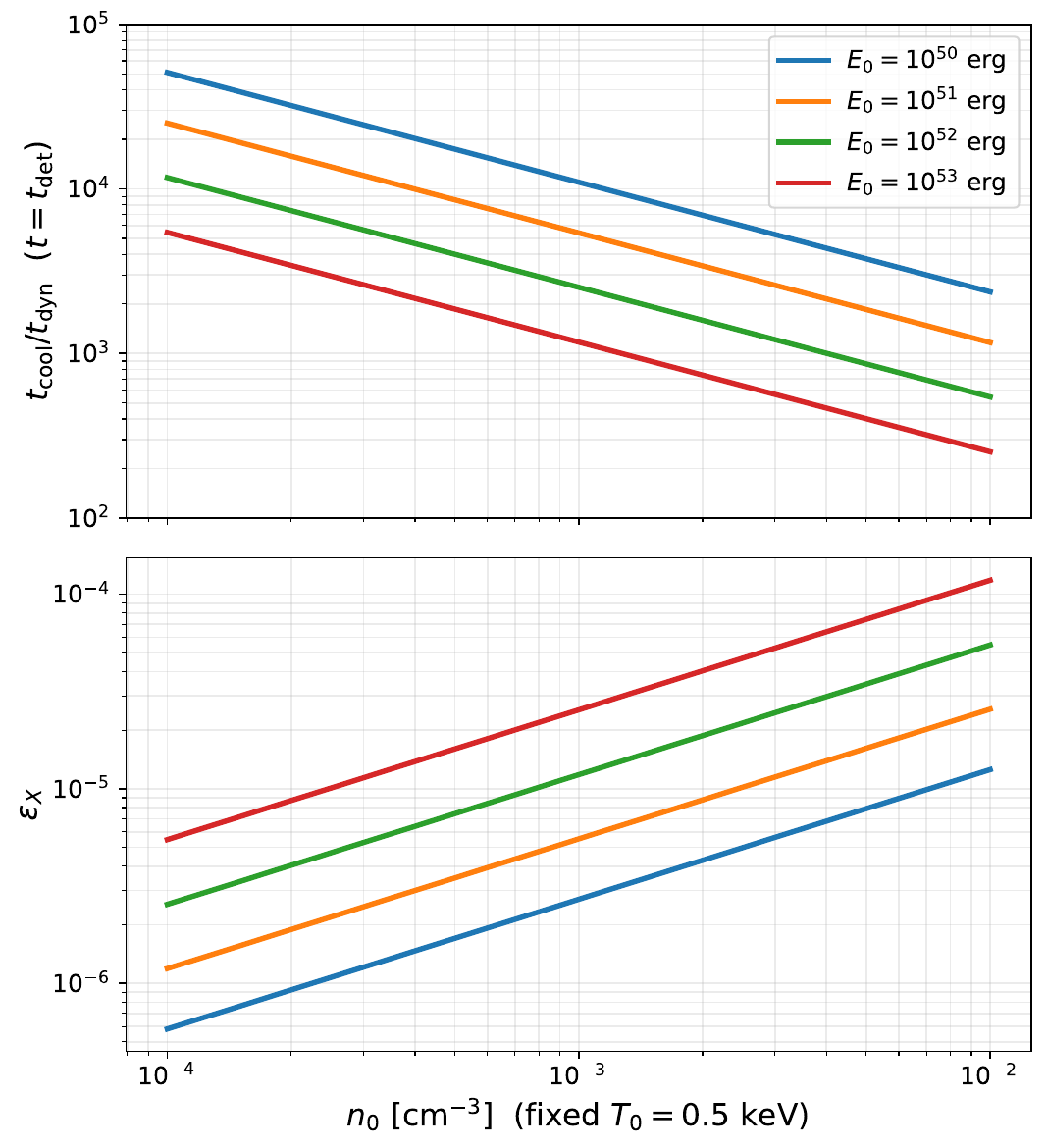}
\caption{Cooling inefficiency and integrated radiative output of expanding bubbles in hot CGM as a function of ambient density $n_0$, for a fiducial temperature $T_0=0.5$~keV. \textit{Top:} The ratio $t_{\rm cool}/t_{\rm dyn}$ evaluated at the detectable-shell epoch $t_{\rm det}$, where $t_{\rm dyn}\equiv R/v$ is obtained from the dynamical solution and $t_{\rm cool}$ is computed using the same optically thin cooling prescription adopted in \S2.7. The large values indicate that radiative losses are dynamically negligible when the shell becomes difficult to detect. \textit{Bottom:} The global radiative efficiency $\epsilon_X$ (Eq.~\ref{eq:RadiativeEfficiencyBubble}), integrated from $t_{\rm min}$ (defined in \S\ref{subsec:numerical_trajectories}) to $t_{\rm stall}$. Colors indicate different explosion energies $E_0$ (as labeled).}\label{fig:global_cooling_efficiency}
\end{figure}

As shown in the lower panel of Fig.~\ref{fig:global_cooling_efficiency}, the baseline hot-CGM models yield $\epsilon_X \ll 1$ over a wide range of explosion energies $E_0$ and ambient densities $n_0$, further confirming that radiative losses are dynamically negligible throughout the lifetime of the bubble. The systematic increase of $\epsilon_X$ with $E_0$ arises because more energetic explosions remain in the pressure-modified expansion regime for longer, reach larger characteristic radii, and maintain temperatures within the X-ray--emissive window over an extended interval, leading to a larger time-integrated X-ray output even though cooling never dominates the energy budget.

Importantly, radiative inefficiency does \emph{not} imply dynamically weak feedback. When cooling is negligible, the injected energy is predominantly redistributed through mechanical channels rather than radiated away. During the supersonic and transonic phases (Stages~II--III), a substantial fraction of the energy is transferred to the CGM via $P\,dV$ work, weak shocks, and compression waves, which irreversibly raise the entropy of the surrounding medium. As the expansion slows and the flow becomes subsonic (Stage~IV), the bubble interior approaches approximate pressure equilibrium with the ambient CGM and no longer drives a shock or sustained expansion. Nevertheless, as the system relaxes toward equilibrium, sound waves and other low-Mach-number compressive disturbances are naturally launched and propagate outward through the hot, low-density CGM, transporting energy to radii comparable to or exceeding $R_{\rm dep}$.

In this late-time regime, the residual thermal and kinetic energy of the bubble is dissipated primarily through a combination of (i) sound-wave damping, (ii) turbulent cascades triggered by shear and buoyancy at the bubble--CGM interface, and (iii) gradual mixing with the ambient medium. While sound waves propagate efficiently through the hot CGM with little dissipation, they can steepen into weak shocks or be damped when encountering cooler or denser gas with lower sound speed, such as cold clumps, filaments, or multiphase interfaces. In this sense, wave-mediated energy transport in a hot CGM is analogous to tsunami propagation in terrestrial oceans, where energy is carried over large distances and released primarily when environmental conditions change. Similar mechanisms of energy transport and dissipation have been widely discussed in the context of AGN feedback and intra-cluster medium (ICM)/CGM heating, where sound waves, weak shocks, and turbulence distribute mechanical energy over large volumes without producing strong radiative signatures \citep[e.g.,][]{Fabian2003,Fabian2006,Forman2007,McNamara2007,Gaspari2012,LiY2014}.

Therefore, even though $\epsilon_X \ll 1$, bubbles expanding into a hot CGM can have a profound and lasting impact on the thermodynamic state of the CGM. The injected energy is not lost, but instead redistributed as elevated entropy and gentle motions over a large volume, suppressing cooling and altering subsequent accretion and recycling processes long after the outward expansion has stalled.

\paragraph{Limitations of the uniform interior assumption and comparison with simulations.}
The analytic framework adopted in this work assumes a thin swept-up shell and a spatially uniform hot interior characterized by volume-averaged density and temperature. While this approximation captures the global dynamical evolution of the bubble, it inevitably simplifies the internal thermodynamic structure. In particular, the X-ray emissivity depends on the emission measure, $\epsilon_X \propto n^2 \Lambda(T)$, and is therefore highly sensitive to density inhomogeneities.

Hydrodynamic simulations of SN explosions in low-density hot media show that the interior structure of the remnant can deviate substantially from the uniform assumption owing to processes such as turbulent mixing, shell corrugation, and entrainment of ambient material. These effects produce localized overdense regions and mixing layers that can significantly enhance $\langle n^2 \rangle$ relative to $\langle n \rangle^2$, thereby boosting the radiative output. Early 1D hydrodynamic simulations already demonstrated that supernova remnants evolving in hot ambient media develop weak shocks and extended low-density interiors, while the blast wave gradually approaches the sound speed of the surrounding medium \citep{Tang2005}. Subsequent studies further explored the dynamical evolution and energy redistribution of supernova remnants in hot environments (e.g., \citealt{Tang2009}).

More recent multidimensional simulations have shown that interactions among supernova remnants and the ambient medium naturally generate multiphase structures and turbulent mixing layers, which can dominate the emission measure and enhance the X-ray luminosity compared to the uniform analytic models developed in this paper \citep{LiM2020a, LiM2020b}. These simulations highlight the importance of density inhomogeneity, mass loading, and turbulent mixing in determining the radiative properties of SN-driven bubbles.

Consequently, the X-ray radiation efficiency predicted by the present analytic framework should be regarded as a lower limit. The simplified treatment does not include density inhomogeneity, turbulent mixing, or mass loading, all of which can increase the emission measure and enhance the radiative output in realistic environments. Observational estimates of the X-ray radiation efficiency of SN heated hot CGM are typically of order $\sim$1\% (e.g., \citealt{LiJ2011,LiJ2013b,LiJ2017}), while numerical simulations generally predict values intermediate between analytic estimates and observations (typically $\sim10^{-4}-10^{-3}$; e.g., \citealt{LiM2020a,LiM2020b}). The discrepancy mainly reflects the strong dependence of radiative emission on density structure, which is not resolved in the present analytic description.

%---------------------------------------------------------------
\subsection{Ionic columns and absorption signatures}
\label{subsec:ionic_columns}
%---------------------------------------------------------------

High-ionization absorption lines provide a complementary and often more sensitive probe of feedback-driven disturbances in the CGM, particularly in regimes where diffuse X-ray emission is weak and falls below current surface-brightness limits. Over the past decade, both UV and X-ray absorption-line observations have established the widespread presence of highly ionized metals in galactic halos, including \ion{O}{6}, \ion{O}{7}, and \ion{O}{8}, tracing gas spanning temperatures from $T\sim10^{5.5}$~K to $\gtrsim10^{6.5}$~K \citep[e.g.,][]{Tumlinson2011,Nicastro2016,Bregman2018}.

X-ray absorption detections of the hot CGM are observationally challenging and typically require relatively large ionic column densities. For current-generation instruments, individual detections generally correspond to
$N_{\rm OVII}\gtrsim10^{15}$--$10^{16}\,\mathrm{cm^{-2}}$ and
$N_{\rm OVIII}\gtrsim\mathrm{few}\times10^{15}\,\mathrm{cm^{-2}}$,
with exact thresholds depending on spectral resolution, signal-to-noise ratio, and line saturation effects \citep[e.g.,][]{Nicastro2018}. As a result, most constraints on hot CGM absorption come from stacked samples or from carefully selected sightlines toward bright background AGNs.

\begin{figure*}
  \centering
    \includegraphics[width=0.49\textwidth]{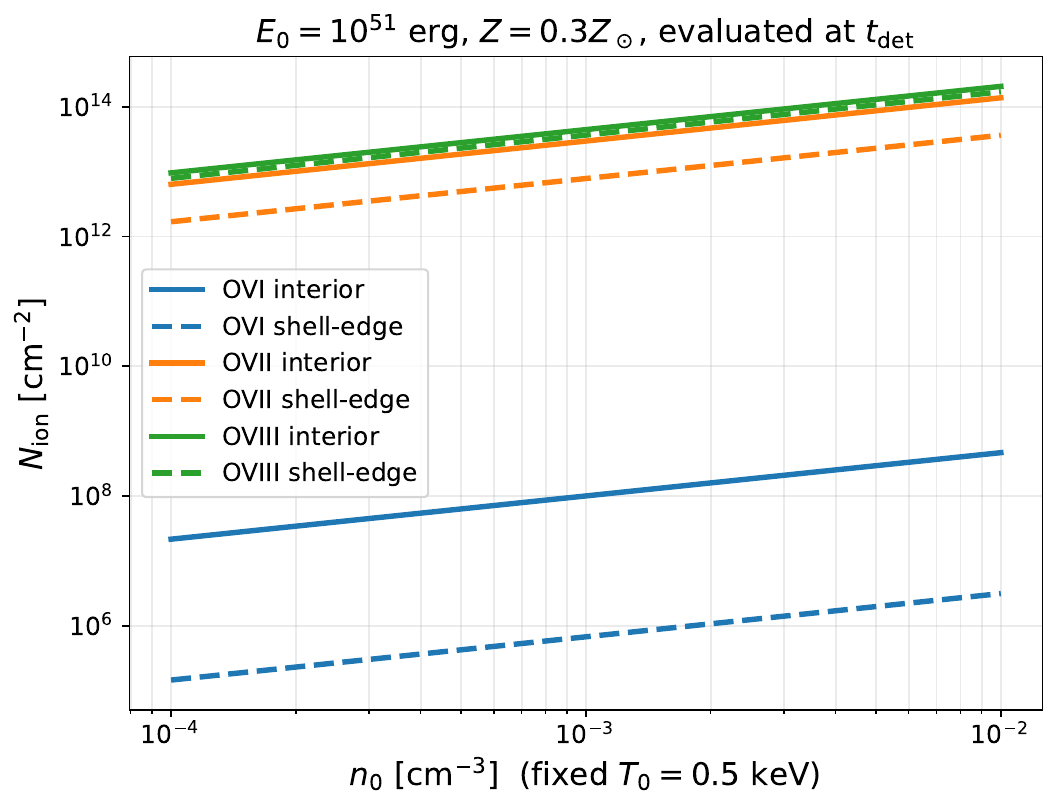}
    \includegraphics[width=0.49\textwidth]{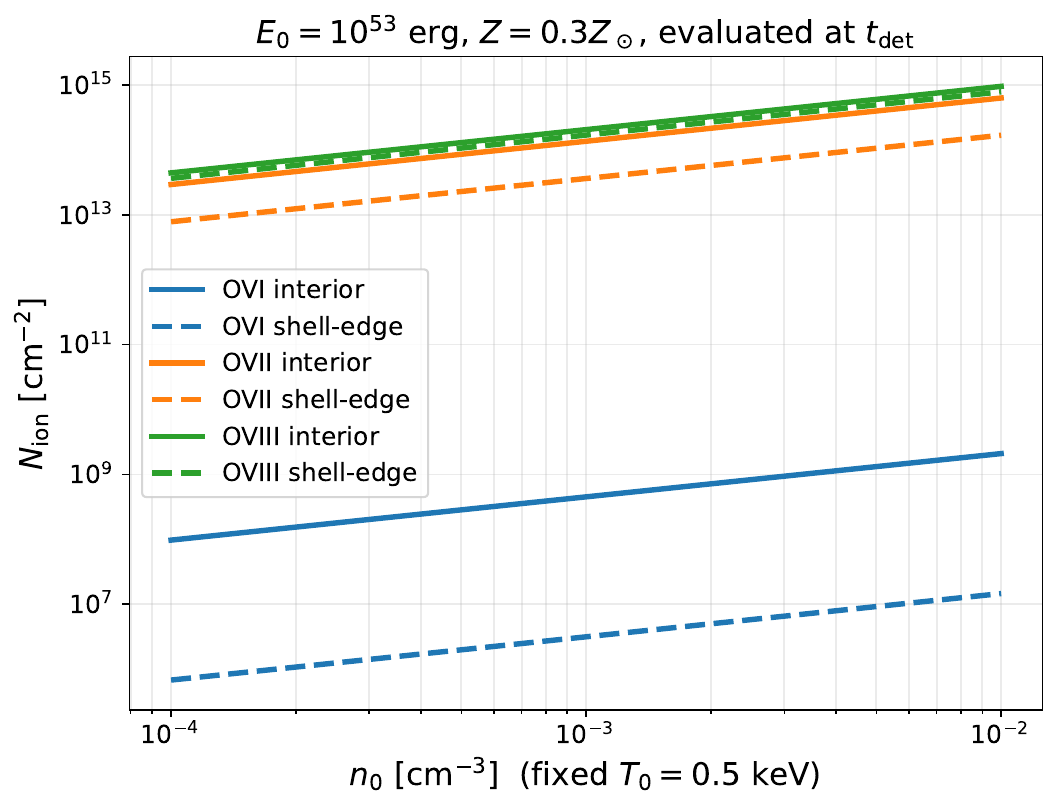}
  \caption{Reference oxygen ionic columns at the detectable-shell epoch $t_{\rm det}$ for a single-event bubble in a hot CGM, shown as a function of ambient density $n_0$ at fixed $T_0=0.5$~keV and $Z=0.3\,Z_\odot$. Each panel plots $N_{\rm ion}$ for \ion{O}{6}, \ion{O}{7}, and \ion{O}{8} for two representative sightlines: an ``interior'' chord through the bubble interior (solid) and a ``shell-edge'' chord passing close to the outer shell (dashed). \textit{Left:} $E_0=10^{51}$~erg. \textit{Right:} $E_0=10^{53}$~erg.}\label{fig:global_ion_columns}
\end{figure*}

Fig.~\ref{fig:global_ion_columns} shows the predicted column densities of representative high-ionization species as a function of ambient density $n_0$ and injected energy $E_0$, evaluated at $t_{\rm det}$ defined in \S\ref{subsubsec:Rdet_tdet}. Several systematic trends are evident. First, for fixed CGM conditions, larger $E_0$ leads to higher ionic columns, reflecting the larger volume of gas that is shock-heated or dynamically perturbed by more energetic events. Second, different ions preferentially trace different temperature regimes within the disturbed region: \ion{O}{6} is most sensitive to intermediate-temperature gas near the peak of the cooling curve, while \ion{O}{7} and \ion{O}{8} predominantly arise from hotter, more volume-filling gas in the bubble interior and post-shock regions. $N_{\rm OVI}$ is a few orders of magnitude lower than $N_{\rm OVII}$ and $N_{\rm OVIII}$, mainly because the temperature of the gas in the bubble interior is high enough (Fig.~\ref{fig:Tbar}).

For fiducial hot-CGM densities, the predicted \ion{O}{7} and \ion{O}{8} column densities approach current X-ray absorption detection thresholds only for the most energetic events (right panel of Fig.~\ref{fig:global_ion_columns}; assuming $E_0=10^{53}\rm~erg$). Single SN energy injection typically produces columns below the level required for individual detections, consistent with the observational difficulty of identifying hot CGM absorption associated with isolated feedback episodes. This suggests that observable X-ray absorption is more naturally associated with either energetic outbursts or the cumulative effect of repeated or overlapping events along the line of sight.

Because the bubble continuously sweeps up ambient material as it expands, the total column associated with the swept-up gas generally increases toward the late stages of the evolution, even though the density inside the bubble decreases with time. For this reason, the ionic columns shown in Fig.~\ref{fig:global_ion_columns} are evaluated at $t_{\rm stall}$, which roughly corresponds to the largest spatial extent reached by the expanding structure so represents the largest column densities a bubble can reach. In practice, however, the observable ionic columns depend much more sensitively on the uniformity of the bubble interior, the thermal and ionization state of the gas than on the total swept-up mass alone.

Similar as above, because the model considered here describes a single, instantaneous energy injection in a smooth ambient medium, the resulting column densities should be regarded as reference levels rather than direct predictions for observed CGM absorbers. In realistic galaxies, clustered supernovae, continuous energy injection, environmental inhomogeneity, and line-of-sight superposition can substantially enhance the effective path length and time-averaged ionic columns. Within this context, the expanding bubble model provides a physically motivated baseline linking feedback energetics to CGM absorption signatures.

%===============================================================
\section{Model extensions and qualitative comparison with observations}
\label{sec:discussion}
%===============================================================

The baseline model developed in \S\ref{sec:baselinemodel} adopts a deliberately controlled setup: a single, instantaneous energy injection expanding into a uniform hot CGM. This choice isolates the dynamical role of finite ambient pressure and sound speed, and leads to the central result that, in hot halos, feedback-driven bubbles are typically radiatively inefficient and terminate through pressure confinement and/or transonic relaxation rather than classical radiative shell formation (as also claimed in \citealt{Churazov2001,Tang2005,Tang2009}).

In this section we discuss physically motivated extensions of the baseline model and compare them qualitatively with observations. Our aim is not to construct a new quantitative solution, but to clarify which assumptions must be relaxed to produce signatures comparable to observations and how the ``hidden energy'' identified in \S\ref{sec:global_outcomes} may become observationally visible.

%---------------------------------------------------------------
\subsection{Continuous energy injection and AGN-driven bubbles}
\label{subsec:AGN_continuous}
%---------------------------------------------------------------

A natural extension of an instantaneous SN-like injection is continuous or episodic energy input, as expected for AGN-driven bubbles or nuclear outflows \citep{Weaver1977,Begelman1989,Faucher2012}. A key physical distinction between SN and AGN feedback lies in the ratio of injected energy to injected mass, or the specific energy $\mathcal{E}$. In both cases of SNe and AGN feedback, the ultimate energy source is gravitational potential, and the characteristic specific energy may be written generically as:
\begin{equation}\label{eq:SpecificEnergy}
\mathcal{E}_{\rm grav} \sim \frac{GM}{R},
\end{equation}
where $M$ and $R$ denote the mass and characteristic radius of the gravitating object from which the energy is extracted. Expressed in this form, the corresponding energy per particle (or temperature scale) follows directly from $kT \sim \mu m_p \mathcal{E}_{\rm grav}$. For a very rough estimation, adopting the typical mass and radius of a compact star (white dwarf or neutron star) and a supermassive black hole (SMBH), the ratio of the specific energy of a SN and an AGN is typically in the range of $\mathcal{E}_{\rm AGN}/\mathcal{E}_{\rm SN}\sim10^{2-4}$, not considering the energy conversion efficiencies. This ensures a much higher specific energy of AGN than SNe feedback.

This large contrast in specific energy has direct dynamical consequences for how energy injection proceeds. In SN feedback, the injected energy is inseparably linked to a substantial ejecta mass, making the process effectively \emph{mass-limited}: increasing the injected energy necessarily implies increasing the injected mass. As a result, the early evolution is governed by a well-defined free-expansion phase, and the swept-up timescale $t_{\rm sw}$ plays a central role in setting the subsequent dynamics. In contrast, AGN feedback is characterized by extremely high specific energy, such that only a negligible amount of mass needs to be coupled to the outflow in order to deliver a given mechanical power. The injection is therefore \emph{power-limited}, regulated primarily by the accretion rate and efficiency of the central engine rather than by mass loading. In this regime, $t_{\rm sw}$ becomes formally very short and dynamically unimportant, and the bubble evolution is instead controlled by pressure balance, sound-crossing, and transonic transition timescales. This decoupling between energy and mass injection allows AGN to sustain mechanical feedback over extended periods, either continuously or through repeated episodes, in contrast to SN feedback where sustained activity inevitably accumulates mass and reduces the effective specific energy (so increase the radiative cooling) of the bubble.

Consistent with this picture, another key difference between SNe and AGN feedback lies in the duration of energy injection: SNe deposit their energy impulsively on short dynamical timescales, while AGN feedback can inject energy over much longer intervals, either continuously or through discrete episodes. To connect AGN-driven feedback to the baseline bubble evolution framework as discussed in \S\ref{sec:baselinemodel}, we treat a single AGN activity episode of duration $t_{\rm inj}$ and characteristic mechanical power $L_{\rm AGN}$ as depositing a total energy:
\begin{equation}\label{eq:AGNE0}
E_0 \equiv E_{\rm AGN} \simeq L_{\rm AGN}\,t_{\rm inj}.
\end{equation}
Observationally inferred mechanical powers of AGN outbursts from X-ray cavity and shock measurements span $L_{\rm AGN}\sim10^{43}$--$10^{45}\,{\rm erg\,s^{-1}}$ \citep[e.g.,][]{Birzan2004,Birzan2008,McNamara2007,Fabian2012}. Typical durations inferred from cavity ages, buoyant rise times, and AGN duty cycle arguments are $t_{\rm inj}\sim10^{6}$--$10^{8}$~yr \citep[e.g.,][]{Churazov2001,Hopkins2006,Novak2011}, implying total injected energies $E_0\sim10^{56}$--$10^{60}$~erg, consistent with direct cavity enthalpy estimates in groups and clusters \citep[e.g.,][]{McNamara2005,McNamara2012}.

In the discussion below we evaluate the corresponding characteristic timescales using the baseline scalings in Eqs.~\ref{eq:tswnorm}, \ref{eq:tcoolnorm}, \ref{eq:tPnorm}, \ref{eq:tMnorm}, adopting fiducial hot-CGM conditions $n_0=10^{-3}\,{\rm cm^{-3}}$ and $T_0=0.5$~keV unless otherwise noted. For these parameters, the pressure-equilibrium time and the transonic transition time scale as: $t_P,t_{\mathcal{M}}\propto E_0^{1/3}n_0^{-1/3}T_0^{-5/6}$. Numerically, this yields:
\begin{align}\label{eq:tPtMAGN}
t_P &\simeq (21,\ 45,\ 96,\ 207,\ 446)\ {\rm Myr},\\
t_{\mathcal{M}} &\simeq (32,\ 69,\ 149,\ 322,\ 693)\ {\rm Myr},
\end{align}
for $E_0=(10^{56},\,10^{57},\,10^{58},\,10^{59},\,10^{60})$~erg, respectively. The radiative cooling time of the hot bubble interior remains long, $t_{\rm cool}\sim{\rm few}$--$10$~Gyr for $T\sim10^{7}$~K and $n\sim10^{-4}\rm~cm^{-3}$ of a low density AGN blow out bubble interior, consistent with the large $t_{\rm cool}/t_{\rm dyn}$ values shown in Fig.~\ref{fig:global_cooling_efficiency}. The swept-up (free-expansion) timescale $t_{\rm sw}$ is extremely short for such large energies (of order $10^{3}$~yr for SN-like normalizations and scales as $t_{\rm sw}\propto (E_0/10^{51}\,\mathrm{erg})^{-1/2}$) and is typically not a controlling scale for AGN-driven bubbles.

Taken together, these estimates robustly imply
\begin{equation}
t_{\rm cool} \gg t_{\rm inj} \gg t_{\rm sw},
\end{equation}
while the relation between $t_{\rm inj}$ and $t_P$ or $t_{\mathcal{M}}$ depends on the episode energy. For lower-energy or long-duration episodes ($E_0\sim10^{56}$--$10^{57}$~erg, $t_{\rm inj}\sim10$--$100$~Myr), it is plausible that $t_{\rm inj}\sim t_P$ and $t_{\mathcal{M}}$. For more energetic episodes ($E_0\gtrsim10^{58}$~erg), however, one typically finds $t_{\rm inj}<t_P,t_{\mathcal{M}}$ by factors of a few to an order of magnitude.

This ordering has important implications for AGN driven bubble evolution. During the active phase ($t\lesssim t_{\rm inj}$), the bubble is generally still over-pressured with respect to the ambient CGM and expands supersonically (or at least transonically). Because injection typically ceases well before pressure regulation and the transonic transition for high-$E_0$ episodes, most of the subsequent evolution can be approximated as an effectively energy-conserving expansion with total energy $E_0=L_{\rm AGN}t_{\rm inj}$, followed only later by the pressure-modified and subsonic continuation. Scaling from Fig.~\ref{fig:global_Rmax_lifetime}, such episodes naturally produce bubbles with characteristic sizes of $\sim10$--$100$~kpc in hot halos, comparable to the cavities commonly detected in galaxy groups and clusters (e.g., \citealt{Fabian2003,Fabian2006,Birzan2004}).

The observational detectability of these bubbles, however, depends sensitively on the ambient density. In clusters and massive groups, the higher CGM/ICM density yields a strong X-ray surface-brightness contrast between a cavity and the surrounding gas, making such bubbles readily detectable (e.g., \citealt{Fabian2003,Fabian2006}). Around individual field galaxies, the hot CGM density is typically lower, and the X-ray emissivity scales roughly as $n^2$, so even bubbles of comparable physical size can have very low surface-brightness contrast and be difficult to detect (e.g., \citealt{LiJ2008,Heald2022}). In these environments, identifying AGN-driven bubbles often requires additional mechanisms that enhance X-ray emission or contrast, such as compression, mixing, or localized cooling (e.g., \citealt{LiJ2019,LiJ2022,LiJ2024a}), which we discuss in the following subsections.

%---------------------------------------------------------------
\subsection{Mechanisms to enhance the X-ray radiation efficiency}
\label{subsec:xray_enhancement}
%---------------------------------------------------------------

As demonstrated in \S\ref{sec:global_outcomes}, low X-ray radiation efficiency does not imply weak feedback. When cooling is inefficient, injected energy is stored primarily in thermal pressure and bulk motions and redistributed through mechanical channels such as weak shocks, sound waves, and gradual mixing into the ambient CGM \citep{Churazov2001,Tang2009}. The baseline model of bubble explosion in the hot CGM explored in \S\ref{sec:baselinemodel} and \ref{sec:global_outcomes} represents this mechanically efficient but X-ray–faint limit.

The X-ray emissivity scales as $j_X\propto n^2 \Lambda_X(T)$, so enhancing the observable X-ray output requires either increasing the gas density, shifting the temperature into a favorable range, or concentrating energy dissipation into a small volume. In a uniformly hot and tenuous CGM, none of these conditions is naturally satisfied: the bubble interior remains dilute, characteristic temperatures are high, and once the expansion becomes pressure-regulated or transonic, the injected energy is distributed over a large volume rather than dissipated locally, long before radiative cooling can dominate (\S\ref{sec:global_outcomes}; \citealt{Tang2005}). Several physical mechanisms can nevertheless increase the X-ray radiation efficiency without changing the total injected energy.

%---------------------------------------------------------------
\subsubsection{Interaction with colder and/or denser gas}
\label{subsubsec:ColdHotInteraction}
%---------------------------------------------------------------

Coupling between the hot bubble and a colder and/or denser phase --- through cloud crushing, turbulent mixing layers, conductive evaporation, or entrainment --- can raise the density of the X-ray–emitting plasma and shorten the cooling time (e.g., \citealt{McKee1990,Slavin1993,Strickland2002,Armillotta2017}). Mass loading from cool gas is a natural outcome of such multiphase interactions, lowering the effective specific energy of the hot phase and providing an efficient route to enhanced $L_X$ (e.g., \citealt{Strickland2009,LiJ2009,LiJ2011,LiJ2013b}). 

The mass-loading process is intrinsically tied to the spatial distribution and fragmentation of cold cloudlets, and the resulting density enhancement is highly inhomogeneous. Because the X-ray emissivity scales as $j_X \propto n^2$, the increase in X-ray luminosity is therefore dominated by localized high density interfaces rather than being proportional to the total mass of cold gas mixed into the hot phase. In typical X-ray observations, limited angular resolution and photon statistics often prevent these fine structures from being resolved (with some possible exceptions, e.g., \citealt{LiJ2008,LiJ2024a}), so the inferred $L_X$ may not provide a direct or reliable measure of the total mass-loading rate.

%---------------------------------------------------------------
\subsubsection{Localized or anisotropic energy deposition}
\label{subsubsec:AnisotropicEnergyDeposition}
%---------------------------------------------------------------

Energy injection that is spatially concentrated --- such as collimated AGN jets or over-pressured bipolar nuclear outflows --- can deposit energy and momentum into the ambient hot gas on timescales much shorter than those required for pressure equilibration or global expansion (e.g., \citealt{Heinz1998,Krause2012,LiJ2022}). In this regime, the outflow behaves more like a ballistic ``bullet'' than a pressure-driven balloon: dissipation occurs locally and impulsively, and the shocked gas cannot immediately re-expand (e.g., \citealt{Heinz2006,Morsony2010,Cielo2018}). As a result, compression is maintained over a finite residence time rather than being rapidly diluted by adiabatic expansion.

Pure hydrodynamic Rankine--Hugoniot jump conditions limit the instantaneous density contrast of a strong shock to a factor of $\sim4$, which by itself is often insufficient to account for the brightest compact X-ray features in a hot ambient medium. However, several additional physical processes can further enhance the effective compression ratio in such localized interactions. Radiative post-shock cooling can reduce pressure support and allow continued compression (e.g., \citealt{Dopita1996,Allen2008}); magnetic draping and tension can suppress lateral expansion and promote density buildup (e.g., \citealt{Lyutikov2006,Dursi2008,LiJ2024a}); and cosmic-ray--modified shocks can achieve compression ratios significantly exceeding the hydrodynamic limit (e.g., \citealt{Drury1981,Berezhko1999,Guo2008,LiJ2019}). Repeated shock passages and shock focusing in confined geometries can further amplify density contrasts (e.g., \citealt{Heinz2006,Cielo2018}). Together, these effects can produce density enhancements of order $\sim10$ or more, sufficient for the $n^2$ dependence of the X-ray emissivity to generate strong, localized X-ray emission.

This enhancement arises from localized dissipation and sustained compression of already hot gas, acting independently of --- though potentially in combination with --- the direct cold gas mass loading mechanism discussed above, while the global feedback remains mechanically dominated and radiatively inefficient.

%---------------------------------------------------------------
\subsubsection{Temporal intermittency of energy injection}
\label{subsubsec:IntermittencyInjection}
%---------------------------------------------------------------

Even for a fixed time-averaged mechanical power, the temporal structure of energy injection can significantly affect how efficiently injected energy is converted into long-lived thermal energy and observable X-ray emission. Compared to smooth or quasi-continuous AGN injection, strongly intermittent feedback deposits energy in discrete bursts, each with a finite energy and duration. Although the instantaneous power during an individual burst may exceed that of a continuous injection with the same time-averaged luminosity, a larger fraction of the injected energy is dissipated or lost during each burst --- through shock heating, $PdV$ work, wave excitation, and partial expansion --- before it can accumulate as persistent thermal energy in the bubble interior. As a result, intermittent injection limits the monotonic buildup of entropy and suppresses extreme peak temperatures, allowing a larger fraction of the gas to remain in or repeatedly pass through temperature ranges where the X-ray cooling function $\Lambda_X(T)$ is relatively efficient. This increases the time-integrated X-ray emissivity for a given total injected energy, compared to a single, long-lived episode or a smooth continuous injection \citep{Gaspari2012}.

Such intermittency is both theoretically expected and observationally supported on timescales relevant to CGM-scale dynamics. X-ray cavity ages, buoyant rise times, and multiple generations of cavities and shocks indicate that AGN feedback commonly proceeds through discrete episodes with characteristic durations and separations of $\sim10$--$100$~Myr in galaxy groups and clusters (e.g., \citealt{Birzan2004,McNamara2007,Fabian2012}). Variability on much shorter timescales, such as year-scale AGN flickering or tidal disruption events (TDEs), is unlikely to have a significant impact on the large-scale hot halo, as such fluctuations are rapidly averaged out over the sound-crossing and expansion times of the CGM. In this regime, the degree of intermittency --- characterized by the burst energy relative to the total injected energy and by the separation between successive episodes --- controls how efficiently injected energy is retained as long-lived thermal energy versus dissipated through shocks, $PdV$ work, and waves. This, in turn, regulates the fraction of feedback energy that becomes observable in X-rays, even when the global feedback remains mechanically dominated and radiatively inefficient.

%---------------------------------------------------------------
\subsubsection{External confinement}
\label{subsubsec:ExternalConfinement}
%---------------------------------------------------------------

Confinement by external forces can substantially modify the thermodynamic evolution of feedback-heated gas by inhibiting adiabatic expansion, thereby maintaining higher densities and enhancing radiative losses. In practice, three related but physically distinct forms of confinement may operate: gravitational confinement by the galactic potential, confinement by a high-pressure ambient medium, and magnetic confinement by ordered or tangled fields.

\emph{Gravitational confinement} arises when feedback-heated gas expands within a deep potential well, such that gravity limits buoyant rise and large-scale expansion. In galactic nuclei, massive bulges, and group-scale halos, the gravitational restoring force can prolong the residence time of hot gas at moderate densities and delay dilution by expansion, thereby enhancing radiative losses relative to the freely expanding bubble case. This effect is commonly invoked in models and observations of AGN feedback and cooling flow regulation in massive galaxies, groups and clusters (e.g., \citealt{Churazov2001,McNamara2007,Fabian2012,Yang2016,LiJ2017}).

\emph{Ambient pressure confinement} is instead controlled by the external thermal pressure of the surrounding medium. In high-density environments such as galaxy clusters and rich groups, the elevated ICM pressure can strongly inhibit adiabatic expansion of feedback-driven bubbles, maintaining high interior densities and boosting X-ray emissivity. This mechanism is most effective where the background pressure approaches or exceeds the internal pressure of the outflowing gas and has been invoked to explain the enhanced X-ray emission of galaxies in dense environments (e.g., \citealt{LiJ2013b,LiJ2014}).

\emph{Magnetic confinement} operates through magnetic pressure and tension, particularly when magnetic field lines are stretched, amplified, or draped around rising bubbles or shock fronts. Magnetic stresses can suppress lateral expansion, reduce mixing, and stabilize dense structures against disruption, effectively increasing gas density and enhancing X-ray emissivity. Such effects have been demonstrated in magnetohydrodynamic (MHD) simulations of AGN-inflated bubbles and CR-supported outflows (e.g., \citealt{Ruszkowski2007,Lyutikov2006,Dursi2008,Ruszkowski2017}), and are also supported by observational arguments for pressure equilibrium in nearby galaxies (e.g., \citealt{Irwin2012,Lu2023,LiJ2024a,LiJ2024b}).

These confinement mechanisms are most effective in environments where the gravitational potential, ambient pressure, or magnetic fields provide a confining energy density comparable to that of the expanding bubble interior --- such as galactic centers and group- or cluster-scale halos --- and are generally subdominant in low-mass field galaxies with diffuse hot halos (e.g., \citealt{LiJ2013b,LiJ2017}). In such confined regimes, feedback can remain mechanically dominated on global scales while exhibiting locally enhanced X-ray radiative efficiency.

%---------------------------------------------------------------
\subsubsection{Energy partition into non-thermal channels}
\label{subsubsec:NonthermalChannels}
%---------------------------------------------------------------

As discussed above, feedback energy need not be deposited primarily in the thermal plasma. If a significant fraction of the injected energy is carried by CRs, magnetic fields, or turbulence, the fraction of energy thermalized into the hot plasma can be reduced, suppressing the global thermal X-ray luminosity relative to the total feedback energy (e.g., \citealt{Guo2008,Ruszkowski2017}), even though local confinement or compression due to the lower specific energy may still enhance X-ray emission in specific regions. In such cases, weak X-ray emission reflects a low efficiency of thermal energy conversion rather than a low total feedback strength, consistent with the mechanically efficient but X-ray faint regime emphasized throughout this paper.

It is worth emphasizing that, although charge exchange may act as an important non-thermal emission mechanism that enhances specific soft X-ray emission lines at interfaces between hot and cold gas (e.g., \citealt{Liu2012,Zhang2014}), it is unlikely to dominate the global X-ray energetics in the hot, volume-filling CGM considered here.

%---------------------------------------------------------------
\subsubsection{Connection to observed trends}
\label{subsubsec:efficiency_trends}
%---------------------------------------------------------------

Observations of nearby galaxies show that the X-ray radiation efficiency, $\epsilon_X \equiv L_X/\dot E$, is typically small (of order $\sim1\%$), but exhibits systematic variations with mechanical energy input, the availability of cool gas, and the surrounding environment, etc. (e.g., \citealt{Strickland2004,LiJ2011,LiJ2013b,LiJ2017}). At fixed mechanical power, gas-rich systems generally display higher X-ray luminosities, consistent with enhanced mass loading, multiphase interaction, and increased effective gas density, while gas-poor systems tend to remain X-ray faint.

These empirical trends can be interpreted naturally within the framework developed in this subsection. The baseline hot CGM bubble corresponds to a low-$\epsilon_X$ regime in which feedback operates in a volume filling, high sound speed medium. In this limit, low density and rapid expansion suppress radiative losses despite substantial mechanical energy input, and most of the injected energy is stored in thermal pressure and bulk motions rather than being radiated (e.g., \citealt{Churazov2001,Tang2005,Tang2009}). The hot CGM bubble explored in this work therefore represents an idealized, mechanically efficient but intrinsically X-ray–faint limiting case.

Departures from this limit --- through multiphase interaction and mass loading (\S\ref{subsubsec:ColdHotInteraction}), strongly localized or anisotropic dissipation (\S\ref{subsubsec:AnisotropicEnergyDeposition}), temporally intermittent energy injection (\S\ref{subsubsec:IntermittencyInjection}), gravitational, ambient pressure, or magnetic confinement (\S\ref{subsubsec:ExternalConfinement}), and energy partition into non-thermal channels (\S\ref{subsubsec:NonthermalChannels}) --- systematically increase the gas density and/or prolong the residence time of the gas in radiatively efficient states. These processes raise the \emph{observed} X-ray radiation efficiency $\epsilon_X$ as an emergent, large-scale outcome, without requiring changes to the intrinsic energy partition at the feedback launching site. Their relative importance depends on the availability and spatial distribution of cool gas, the geometry and temporal structure of energy injection, and the depth and pressure of the surrounding halo.

From this perspective, the observed joint dependence of X-ray radiation efficiency on energy input, mass loading, and environment does not primarily reflect variations in the total feedback strength, but instead indicates how far a given system departs from the idealized, volume filling, weakly radiative hot CGM limit. Observed X-ray enhancements therefore provide qualitative insight into the microphysics and environment-dependent manifestation of feedback in real galactic halos, including the roles of multiphase structure, confinement, intermittency, and anisotropic dissipation (e.g., \citealt{LiJ2011,LiJ2013b,LiJ2014,LiJ2017,Fabian2012}).

%===============================================================
\section{Summary and Conclusions}\label{sec:summary}
%===============================================================

We have developed an analytic framework for bubbles expanding into a hot CGM, explicitly accounting for finite ambient pressure and sound speed. The key points are as follows.

(1) \emph{Stage structure in a hot ambient medium.} The evolution can be organized into four physically motivated stages: (I) ejecta-dominated free expansion, (II) Sedov--Taylor expansion, (III) pressure-modified evolution approaching pressure balance and/or transonicity, and (IV) post-transonic relaxation. This staging provides a controlled bridge between the early strong-shock phase and the late-time fate of a disturbance in a hot halo.

(2) \emph{Transonicity and pressure balance often end the strong-shock phase before cooling.} In the hot CGM regime emphasized here, radiative losses are frequently dynamically sub-dominant over the bubble lifetime, and the termination of the strong-shock phase is commonly set by the approach to pressure equilibrium and transonic expansion rather than by the formation of a classical radiative shell. Consequently, the late-time evolution produces only a modest spatial and temporal overshoot beyond the pressure-balance/transonic transition before the disturbance stalls and merges with the ambient medium.

(3) \emph{Observable outcomes from a baseline, mechanically efficient but X-ray--faint limit.} Using stitched numerical trajectories anchored by the analytic scalings, we estimate global outcomes including characteristic radii and lifetimes, band-limited X-ray luminosities, and high-ionization ion column densities. The baseline model defines an intrinsically X-ray--faint limiting case appropriate for low-density hot halos, in which low emissivity primarily reflects the combination of low density and rapid expansion rather than weak mechanical feedback.

(4) \emph{Continuous and episodic energy injection.} Extending the framework beyond single-event explosions, we show that continuous or episodic energy injection relevant for AGN-driven bubbles leads to qualitatively different evolution in a hot CGM. In this regime, the bubble can remain over-pressured and transonic for an extended duration, with its characteristic size, lifetime, and energy-deposition radius regulated by the balance between sustained driving and ambient pressure rather than by a single impulsive event. As a result, continuously driven bubbles can maintain large, long-lived cavities over an extended duration. However, the \emph{detectability} of such cavities does not necessarily increase in low-density halos, because X-ray contrast depends strongly on the ambient emissivity ($\propto n_0^2$) and projection. In practice, the most prominent observational examples of AGN-inflated cavities are often found in dense cluster or group atmospheres, where a low-density cavity is easy to identify against a bright background, even if the cavity interior itself is tenuous.

(5) \emph{How real systems become brighter: departures from the hot-CGM limiting case.} We have outlined several mechanisms that can increase observability and the emergent X-ray radiation efficiency, including multiphase interaction and mass loading, localized or anisotropic dissipation, temporally intermittent driving, confinement by gravity/ambient pressure/magnetic stresses, and energy partition into non-thermal channels. These effects raise the effective radiative output largely by increasing gas density and/or prolonging residence times in radiatively efficient states, without requiring changes to the intrinsic energy partition at the launching site.

Overall, this work provides a physically transparent foundation for interpreting X-ray and absorption-line signatures of feedback in hot galactic halos and for extending analytic bubble models beyond the cold-ISM paradigm.

\begin{acknowledgements}
The author thanks the anonymous referee for very helpful and constructive suggestions that significantly improved the manuscript. The author acknowledges the use of an AI-based language model (ChatGPT; OpenAI, San Francisco, USA) for assistance with coding and for improving the clarity and grammar of the manuscript. All scientific ideas, analyses, interpretations, and conclusions presented in this work were conceived and verified by the author, who takes full responsibility for the content.
The author also acknowledges financial support from the National Science Foundation of China (NSFC) through grants 12321003 and 12273111, from the China Manned Space Program through grants CMS-CSST-2025-A04 and CMS-CSST-2025-A10, and from the Jiangsu Innovation and Entrepreneurship Talent Team Program through grant JSSCTD202436.
\end{acknowledgements}

\bibliographystyle{aasjournal}
\bibliography{hot_cgm_superbubble_refs}

\end{document}